\RequirePackage{fix-cm}
\documentclass[natbib,smallextended]{svjour3}       
\smartqed  
\usepackage{graphicx}

\usepackage{amssymb}
\usepackage{amsmath}
\usepackage{bm}
\usepackage{subfigure}
\usepackage{tabularx} 
\usepackage{tabulary}

\newcommand{\eg}{e.g., }

\newcommand{\ie}{i.e., }

\newcommand{\sect}[1]{Section \ref{s:#1}}

\newcommand{\eqn}[1]{Eq.\ (\ref{e:#1})}

\newcommand{\fig}[1]{Fig.\ \ref{fig:#1}}

\newcommand{\Fig}[1]{Figure \ref{fig:#1}}

\newcommand{\tbl}[1]{Table \ref{t:#1}}

\newcommand{\hide}[1]{} 

\graphicspath{{}}

 \journalname{Space Science Reviews}
\begin{document}

\title{Estimations of the seismic pressure noise on Mars determined from Large Eddy Simulations and demonstration of pressure decorrelation techniques for the InSight mission}
\titlerunning{Seismic pressure noise on Mars}      

\author{Naomi Murdoch        \and
        Balthasar Kenda \and
        Taichi Kawamura \and
        Aymeric Spiga \and
        Philippe Lognonn{\'e} \and
        David Mimoun \and
        William B. Banerdt 
}

\institute{N. Murdoch \at
              Institut Sup\'{e}rieur de l'A\'{e}ronautique et de l'Espace (ISAE-SUPAERO), Universit\'{e} de Toulouse, 31055 Toulouse Cedex 4, France \\
              \email{naomi.murdoch@isae.fr}           
                            \and
                             B. Kenda \at
             Institut de Physique du Globe de Paris, Paris, France \\
              \email{kenda@ipgp.fr}           
                \and
              T. Kawamura \at
             Institut de Physique du Globe de Paris, Paris, France \\
              \email{kawamura@ipgp.fr}           
                              \and
                 A. Spiga \at
             Laboratoire de M\'{e}t\'{e}orologie Dynamique, UMR CNRS 8539, Institut Pierre-Simon Laplace, Sorbonne Universit\'{e}s, UPMC Univ. Paris 06, 4 place Jussieu, 75005 Paris, France\\
              \email{aymeric.spiga@upmc.fr}           
                         \and
            P.  Lognonn{\'e}\at
             Institut de Physique du Globe de Paris, Paris, France \\
              \email{lognonne@ipgp.fr}           
                \and
          D. Mimoun \at
              Institut Sup\'{e}rieur de l'A\'{e}ronautique et de l'Espace (ISAE-SUPAERO), Universit\'{e} de Toulouse, 31055 Toulouse Cedex 4, France \\
              \email{david.mimoun@isae.fr}           
              \and
			W. Bruce Banerdt \at
             Jet Propulsion Laboratory, Pasadena, CA  91109, USA \\
                           \email{william.b.banerdt@jpl.nasa.gov}           
           }

\date{Re-Submission Date: 17 February 2017}

\maketitle

\begin{abstract}

The atmospheric pressure fluctuations on Mars induce an elastic response in the ground that creates a ground tilt, detectable as a seismic signal on the InSight seismometer SEIS.  The seismic pressure noise is modeled using Large Eddy Simulations of the wind and surface pressure at the InSight landing site and a Green's function ground deformation approach that is subsequently validated via a detailed comparison with two other methods based on Sorrells' theory \citep{sorrells1971,sorrells1971b}. The horizontal acceleration as a result of the ground tilt due to the LES turbulence-induced pressure fluctuations are found to be typically $\sim$2 - 40 nm/s$^2$ in amplitude, whereas the direct horizontal acceleration is two orders of magnitude smaller and is thus negligible in comparison. The vertical accelerations are found to be $\sim$0.1 - 6 nm/s$^2$ in amplitude.  These are expected to be worst-case estimates for the seismic noise as we use a half-space approximation; the presence at some (shallow) depth of a harder layer would significantly reduce quasi-static displacement and tilt effects.

We show that under calm conditions, a single-pressure measurement is representative of the large-scale pressure field (to a distance of several kilometers), particularly in the prevailing wind direction. However, during windy conditions, small-scale turbulence results in a reduced correlation between the pressure signals, and the single-pressure measurement becomes less representative of the pressure field.   Nonetheless, the correlation between the seismic signal and the pressure signal is found to be higher for the windiest period because the seismic pressure noise reflects the atmospheric structure close to the seismometer. 

In the same way that we reduce the atmospheric seismic signal by making use of a pressure sensor that is part of the InSight APSS (Auxiliary Payload Sensor Suite), we also the use the synthetic noise data obtained from the LES pressure field to demonstrate a decorrelation strategy. We show that our decorrelation approach is efficient, resulting in a reduction by a factor of $\sim$5 is observed in the horizontal tilt noise (in the wind direction) and the vertical noise. This technique can, therefore, be used to remove the pressure signal from the seismic data obtained on Mars during the InSight mission. 

\end{abstract}

\section{Introduction}
The InSight mission, selected under the NASA Discovery program for launch in {2018}, will perform the first comprehensive surface-based geophysical investigation of Mars. The objectives of the InSight mission are to advance our understanding of the formation and evolution of terrestrial planets and to determine the current level of tectonic activity and impact flux on Mars. SEIS (Seismic Experiment for Internal Structures) is the critical instrument for delineating the deep interior structure of Mars, including the thickness and structure of the crust, the composition and structure of the mantle, and the size of the core.  SEIS consists of two independent, 3-axis seismometers: an ultra-sensitive very broad band (VBB) oblique seismometer; and a miniature, short-period (SP) seismometer that provides partial measurement redundancy and extends the high-frequency measurement capability \citep{lognonne2015b}.

\begin{figure}[h!]
\begin{center}
\includegraphics[scale=0.3]{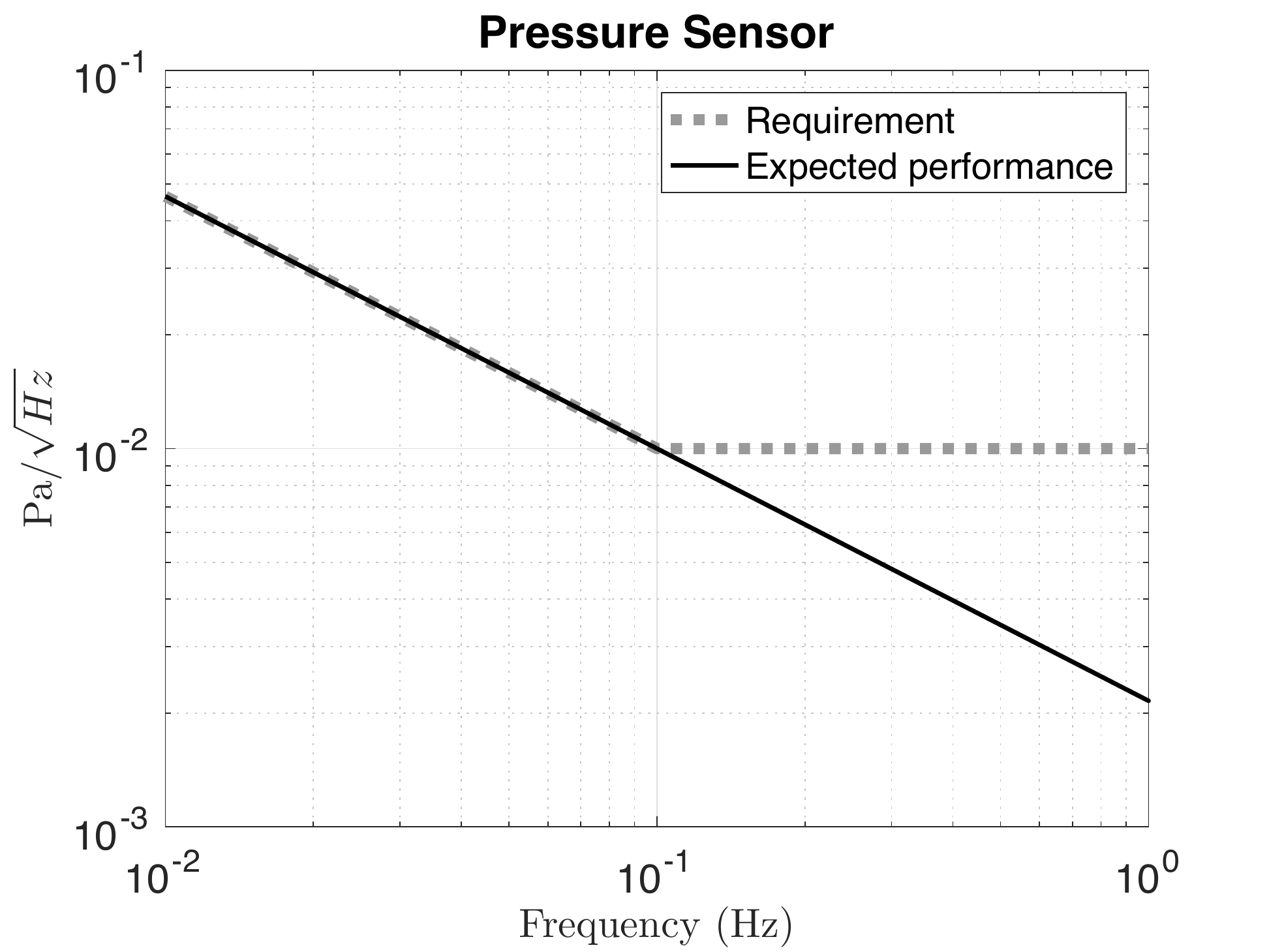}
\caption{InSight pressure sensor sensitivity. The requirement (grey dotted line) and expected performance (black line) of the pressure sensor. }
\label{fig:PressureSensor}
\end{center}
\end{figure}

Meeting the performance requirements of the SEIS instrument is vital to successfully achieve the InSight mission objectives. However, there are many potential sources of noise on seismic instruments. { Also, the different environment on Mars compared to the Earth results in different noise conditions for the Martian seismometer.  Lessons learned from the Viking mission clearly emphasized the importance of protecting the seismometer from Martian meteorological noises \citep{anderson1977}. Almost all the post-Viking Martian seismic observation missions have proposed a wind shield to cover the seismometer in order to reduce the noise from the wind and pressure fluctuation \citep[\eg][]{lognonne96, nishikawa2014}, which is also the case for InSight. Meteorological activities induce noise on the seismometer through various mechanisms, such as the dynamic pressure due to the wind acting directly on the seismometer \citep{lognonne96}, and ground tilt or ground motion due to the interaction of the wind shield or the lander and the Martian winds \citep{murdoch2016, nishikawa2014, lorenz12}. \cite{lognonne96} carried out tests to evaluate the efficiency of a wind shield to protect a seismometer from the wind and showed that, with a wind shield, the noise level can be reduced by a factor of 10 at frequencies lower than 0.05 Hz and by a factor of two at frequencies greater than 1 Hz (the Earth micro-seismic noise makes the noise level estimation difficult between 0.05 and 1 Hz). On the other hand, \cite{nishikawa2014} evaluated a long-period noise due to the ground tilt caused by the torque induced to the wind shield by the wind and suggest the noise to be $10^{-10}$ m/s$^2$/Hz$^{0.5}$ at a windy site and $10^{-9}$ m/s$^2$/Hz$^{0.5}$ at a stormy site in the 1 - 10 mHz bandwidth. This is consistent with \cite{murdoch2016}, who give estimates of $4 \times 10^{-10}$ m/s$^2$/Hz$^{0.5}$ to $1 \times 10^{-9}$m/s$^2$/Hz$^{0.5}$ for the InSight wind shield noise at 10 mHz (assuming a `70\% of the time' wind profile derived from in-situ Phoenix, Viking Lander 1 and Viking Lander 2 data).

Seismometers on Earth are often installed on rigid bedrock in seismic vaults where they are not subjected to the wind, and where the pressure and temperature are stable throughout the observation period. Thus, the atmospheric noise is significantly less compared to the oceanic noise or noise generated from human activities.  Nonetheless, in order to observe small amplitude signals at long periods such as tides or free oscillation of the Earth, such atmospheric noise still needs to be properly treated. The pressure noise on Earth has been studied as a noise source at long-periods of 1-10 mHz, which is below the oceanic micro-seismic bands \citep{zurn1995, beauduin96}. These studies suggests that the pressure noise detected on the seismometers is  $\sim10^{-8}$ m/s$^2$/Hz$^{0.5}$ on the horizontal components and that this noise can be decreased by up to a factor of 10 by applying a correction determined from the correlation between the pressure and seismic data. These pressure sensitivities are, however, related to capsule effects acting on the seismic vault and/or overall tilt acting on the seismic vault, and have therefore a different origin than those acting on seismometers deployed on the surface.

The situation is likely to be more severe on Mars due to the fact that the seismometer will be installed on top of the ground and on a soft regolith layer. Indeed, the temperature variations and the ground tilt due to atmospheric pressure fluctuations are expected to be the major contributors to the seismic noise recorded by the SEIS instrument \citep{mimoun2016}, in addition to non-coherent seismic waves generated by the interaction of the planet's atmosphere with the ground and interior. 
 
The tilt pressure noise has been proposed as the first source of micro-seismic noise by \cite{lognonne93}, and its amplitude has been estimated to be in the range of $10^{-9}-10^{-8}$ m/s$^2$/Hz$^{0.5}$, following the method of \cite{sorrells1971} and \cite{sorrells1971b}, which links the displacement of the ground to the response of a sinusoidal pressure wave.  The temperature noise, with and without windshield, has been estimated with both a field experimental approach by \cite{lognonne98} in the seismic bandwidth and at the longer, tidal period by \cite{VanHoolst2003}. These early estimates have been refined within the SEIS project \citep{lognonne12, mimoun12}, leading to a complete noise model described by \cite{mimoun2016}.

The atmospheric pressure fluctuations on Mars induce an elastic response in the ground that creates ground tilt, vertical displacement, and surface pressure changes. Far from a seismic station, these atmospheric sources excite incoherent seismic waves: at very long periods, global scale circulation and turbulence in the boundary layer will create a background seismic ``hum''  \citep{lognonne2007,nishikawa2017}. Whereas short-term, small-scale (m to 10s of m) atmospheric events such as dust devil episodes will provoke detectable seismic signals both at low frequencies due to ground tilt \citep{lorenz2015} and by the generation of surface waves at higher frequencies. Both of these aspects are studied in detail in \cite{kenda2016}.  

Near, and at, the seismic station, medium-scale atmospheric pressure variations (100s of m to kms) generate ground deformations and, therefore, noise on the seismic records. The investigation of this atmospheric seismic signal is the primary goal of this paper.  In order to allow the detection of smaller amplitude Mars quakes, it is planned to reduce the atmospheric pressure signal by making use of a pressure sensor that will be part of InSight APSS (Auxiliary Payload Sensor Suite). { The requirement and current best estimate of the expected performance of the pressure sensor are given in \fig{PressureSensor}.} Decorrelation techniques will be used to analyze the synchronous pressure and seismic measurements and remove the pressure signal from the seismic signal. The pressure sensor will be on the InSight lander and, thus, almost collocated with the seismometer. } \\

In this paper we aim to answer several important questions:

\begin{itemize}
\item{How representative of the overall pressure field is a single-point measurement of the pressure fluctuations?}
\item{What is the typical amplitude of the tilt signal induced on the seismometer by the pressure variations over a large surface?}
\item{How does this ground tilt correlate with the measurement of a single-pressure sensor at the same location?}
\item{How effective are decorrelation techniques for removing the pressure tilt noise using collocated pressure measurements?}
\end{itemize}

In order to answer these questions, Large Eddy Simulations (LES) of the turbulent fluctuations of wind and surface pressure at the InSight landing site are used. These LES provide us not only with a realistic 2D pressure field, enabling us to compute exactly the surface time-dependent tilts and vertical displacements, but they also provide synthetic wind and pressure data at a given point, which can be used to mimic the APSS recorded data. First, the correlation between the pressure signal at the center of the LES field with the pressure signal in the vicinity is investigated, allowing a characteristic distance of correlation between the pressure signals to be identified. Next, the ground tilt due to atmospheric simulations is modeled by combining the LES and a Green's function static ground deformation model.  The tilt amplitudes are then compared with those predicted by Sorrell's theory \citep{sorrells1971,sorrells1971b}, before considering the correlation of the seismic signal with the collocated pressure signal. Finally, a technique for removing the pressure tilt noise from a seismic signal via decorrelation with collocated pressure measurements is demonstrated { and its performances are given in terms of acceleration ground tilt noise before and after pressure decorrelation}.

\section{The surface environment on Mars}

The Martian environment differs from the terrestrial environment in several respects \citep{InSightERD}. For example, the ambient pressure at the InSight landing site may range from 580 Pa to 1100 Pa (0.5 - 1.0 \% of Earth ambient pressure), the air density is lower (0.015 - 0.025 kg/m$^3$ on Mars compared to the average value of around 1.225 kg/m$^3$ on Earth) and the gravity is smaller (3.71 m/s$^2$ on Mars compared to 9.81 m/s$^2$ on Earth). 
In addition, a layer of granular material (dust and sand) known as regolith covers the red planet.  We assume the ground properties provided in \tbl{regolith} for the Martian regolith, derived from seismic velocity tests of Martian regolith simulant \citep{delage2016}.

 \begin{center}
\begin{table}[h!]
\caption{InSight landing site regolith properties valid for the upper 2 m of regolith measured at a reference pressure of 25 kPa \citep{delage2016}. The error is the standard deviation of the laboratory measurements. }
\begin{tabular}{ccc}
 \hline
 {Bulk density, $\rho_r$} & {S-wave velocity, $v_S$} &  {P-wave velocity, $v_P$} \\  
 {(kg m$^{-3}$)} & {(m s$^{-1}$)} & {(m s$^{-1}$)} \\
 \hline
1665 $\pm$ 38 & 150 $\pm$ 17 & 265 $\pm$ 18 \\
\hline
\end{tabular}
\label{t:regolith}
\end{table}
\end{center}

Currently, the only bodies for which we have quantitative measures of the level of seismicity are the Earth and the Moon. \cite{knapmeyer2006} explain that, with our current knowledge, any model of Mars seismicity cannot be unique, and so an in-situ seismic investigation is necessary to obtain this information. In addition to unknown levels of seismicity, the details of the seismic wave forms on Mars are also currently unknown. This is due to a lack of information about, for example, the source time functions and the wave propagation characteristics and scattering.  However, the amplitude of the Mars seismic signal is expected to be about 4 orders of magnitude lower than on the Earth \citep{lognonne2000}, mainly because of the smaller magnitude quakes expected on Mars.

The investigation of the ground tilt caused by the local pressure field around the seismic station requires the thorough description of the regional pressure field. This is made possible by using turbulence-resolving Large-Eddy Simulations (LES) to describe the atmospheric environment of Mars at the InSight landing site and to model the excitation source, \ie the surface-pressure field. High-resolution LES realistically resolve the Planetary Boundary Layer (PBL) convective motions and the largest turbulent structures such as convective vortices and dust-devils \citep{michaels2004, spiga2009}. These are the main atmospheric features expected to generate long-period seismic noise at the local scale.

\cite{spiga2010} detailed the LES model used in this study; in particular, the physical parametrizations, including radiative transfer, are adapted to the Martian conditions \citep{spiga2009}. The horizontal resolution of the model is 50 m, and the grid covers a region of 14.4 km by 14.4 km. This value is about three times the maximum expected height of the PBL \citep[4.5 km, according to][]{hinson2008}, ensuring the development of convective cells \citep{michaels2004}. In the vertical direction, the grid consists of 151 isobaric levels up to about 8 km; the lowest levels are densely spaced, allowing for a detailed characterization of the interaction with the surface.

For this study, a reference LES was performed with initial and boundary conditions adapted to the 2016 InSight landing site (latitude 4.4$^\circ$N, longitude 136$^\circ$E, altimetry -2652.6 m, albedo 0.26, thermal inertia 260 J m$^{-2}$ K$^{-1}$ s$^{-1/2}$) at the original landing season, Ls=231.2$^\circ$ (northern fall). The delay of the InSight mission to 2018 does not affect the choice of the landing ellipse, and the different landing season (northern spring, L$_s$=19$^\circ$) has a minor impact because of the very low latitude of the InSight landing site. The simulation starts at 8 am local time, and the vertical temperature profile is initialized according to the predictions of the Mars Climate Database \citep{millour2015}. With an output every 6 seconds, the simulation lasts until 9 pm local time, and thus covers the development and the collapse of the PBL convection as well as part of  the calm nighttime period. Moreover, a West-to-East ``background'' horizontal wind of 10 m/s mimics the effects of regional-scale circulation and advects convective cells and vortices towards the East. The direction of the background wind is chosen arbitrarily.

Diurnal variations of pressure associated with global-scale atmospheric thermal tides are not included in LES computations because their typical timescales are of $\sim$2h, which are much larger than the observing sequences of SEIS in which any decorrelation will be needed \citep[the ``hum'' caused by global-scale circulations is discussed in][]{nishikawa2017}.

\begin{figure}[h!]
\begin{center}
\includegraphics[scale=0.3]{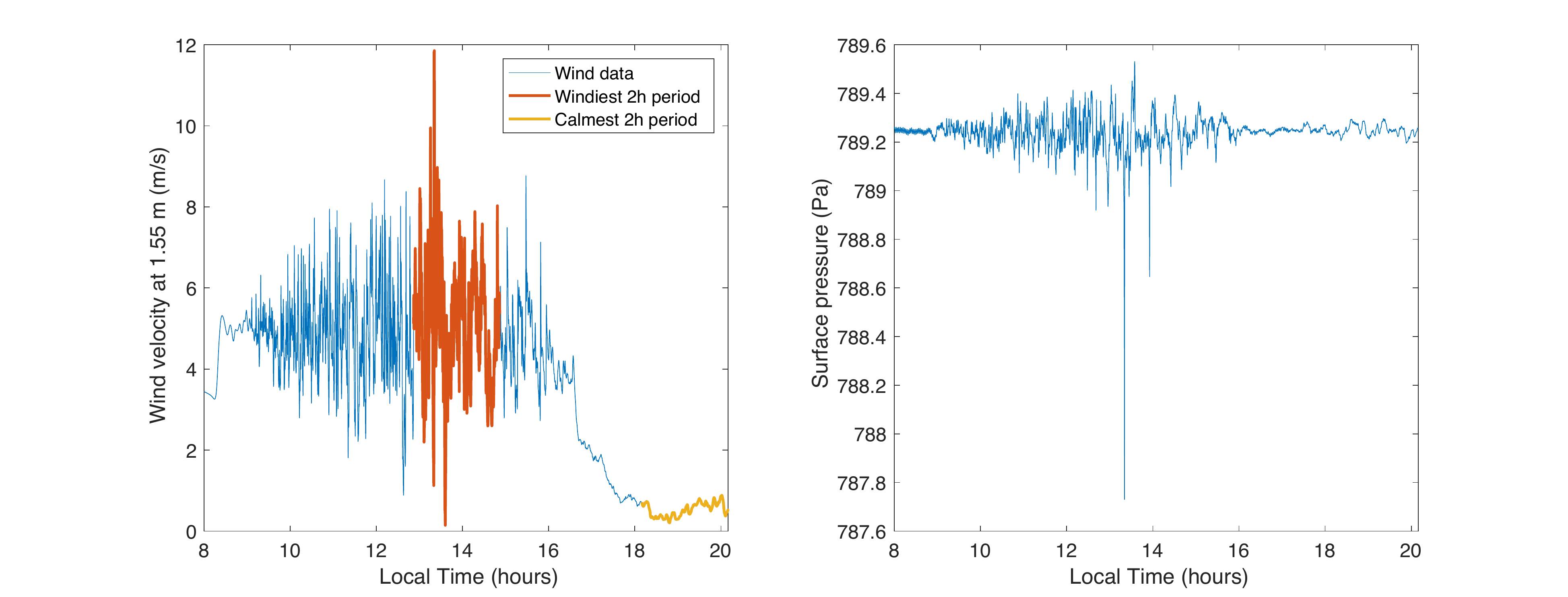}
\caption{LES wind velocity and pressure -  (Left) Horizontal wind velocity at a height of 1.55 m above the surface as a function of time in the center of the LES grid. The wind velocity is calculated from the friction velocity assuming a surface roughness of 1 cm. The windiest period (from 12.9h to 14.9h) is shown by the orange line, and the calmest period (from 18.2h to 20.2h) is indicated in yellow. (Right) Surface pressure as a function of time in the center of the LES grid. The start time in both of these figures is 8 am local time.}
\label{fig:WindPressurecenterOfGrid}
\end{center}
\end{figure}

A snapshot of the LES is shown in \fig{PressureForceExample}. Wind velocity and surface pressure at the center of the grid are shown in \fig{WindPressurecenterOfGrid} for the whole duration of the LES. The wind is computed at a height of 1.55 m; due to the logarithmic profile of the wind velocity in the vertical direction (obtained with a surface roughness of 1 cm), the mean value at 1.55 m is slightly lower than the imposed background wind of 10 m/s. The fluctuations in the two time series are induced by turbulent convective phenomena that are present particularly between 11 am and 4 pm local time. The large pressure drops seen in \fig{WindPressurecenterOfGrid} are due to convective vortices \citep[see further discussion in \sect{seismic} and][]{kenda2016}.

\section{Pressure correlation considerations}
\label{s:pressurecorrelation}

Here the correlation between the pressure measurements at the center of the grid with pressure measurements in the vicinity is investigated, allowing a characteristic correlation length between the pressure measurements to be identified.  This provides an understanding of how representative a single-point measurement of the pressure fluctuations is of the overall pressure field.

\begin{figure}[!h]
\begin{center}
\includegraphics[scale=0.15]{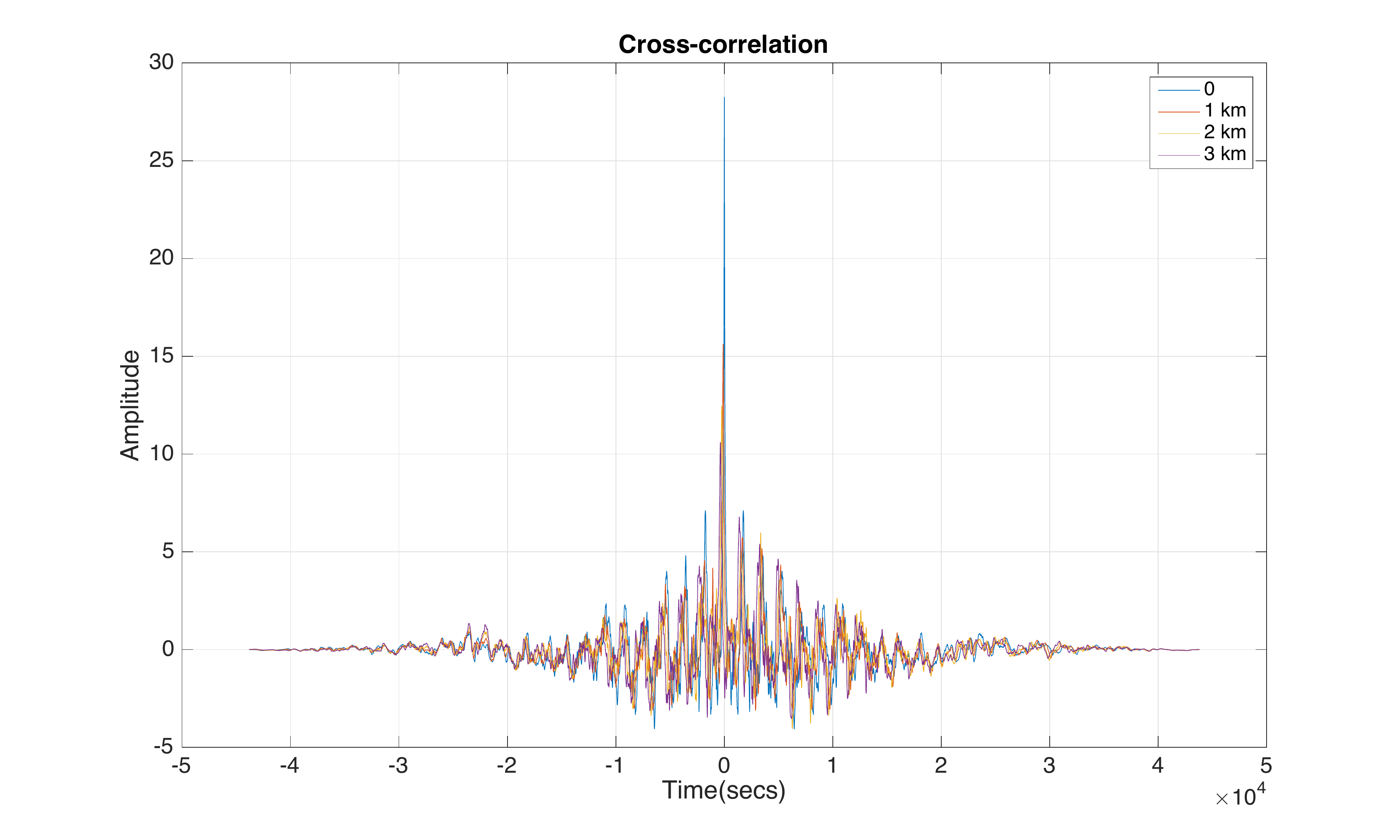}
\includegraphics[scale=0.15]{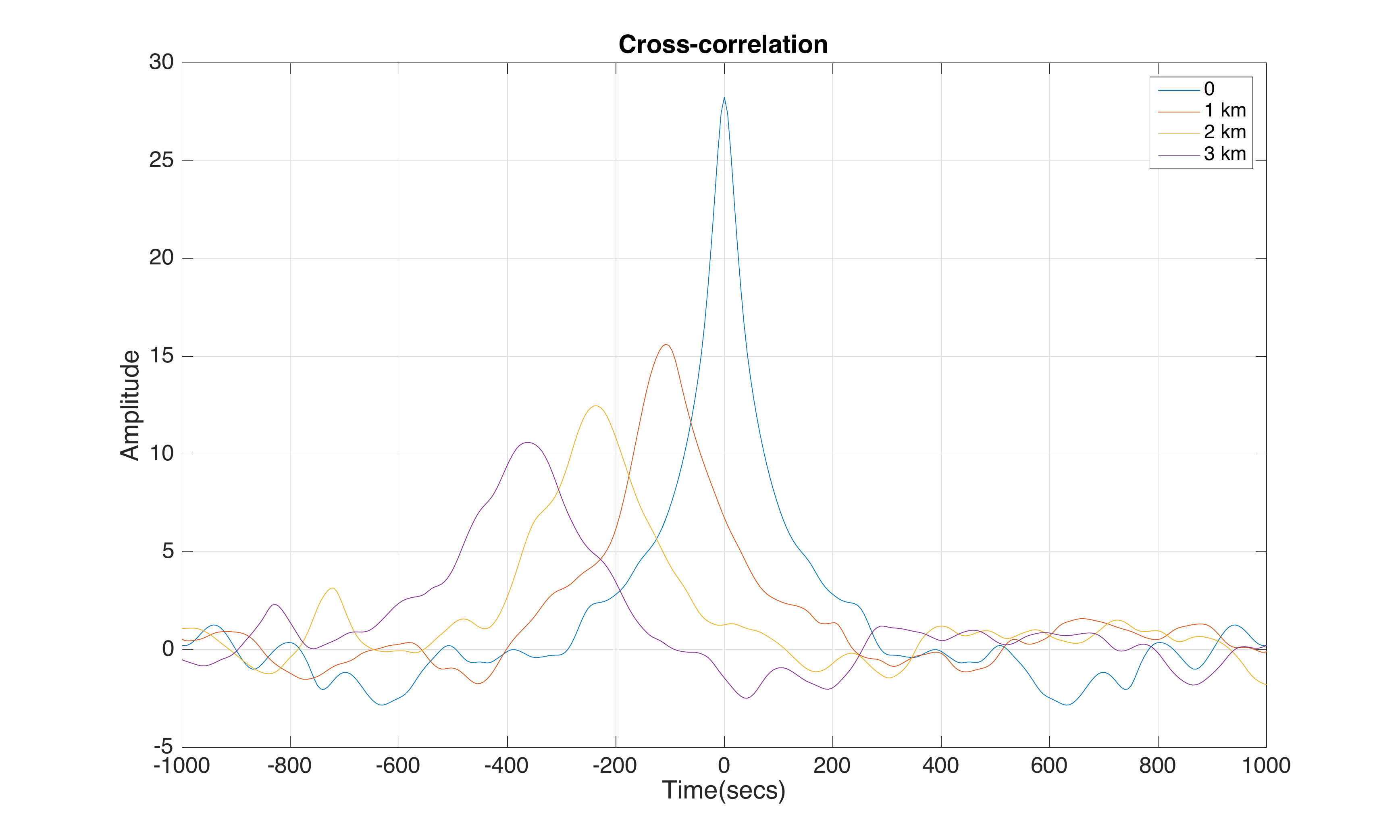}
\includegraphics[scale=0.15]{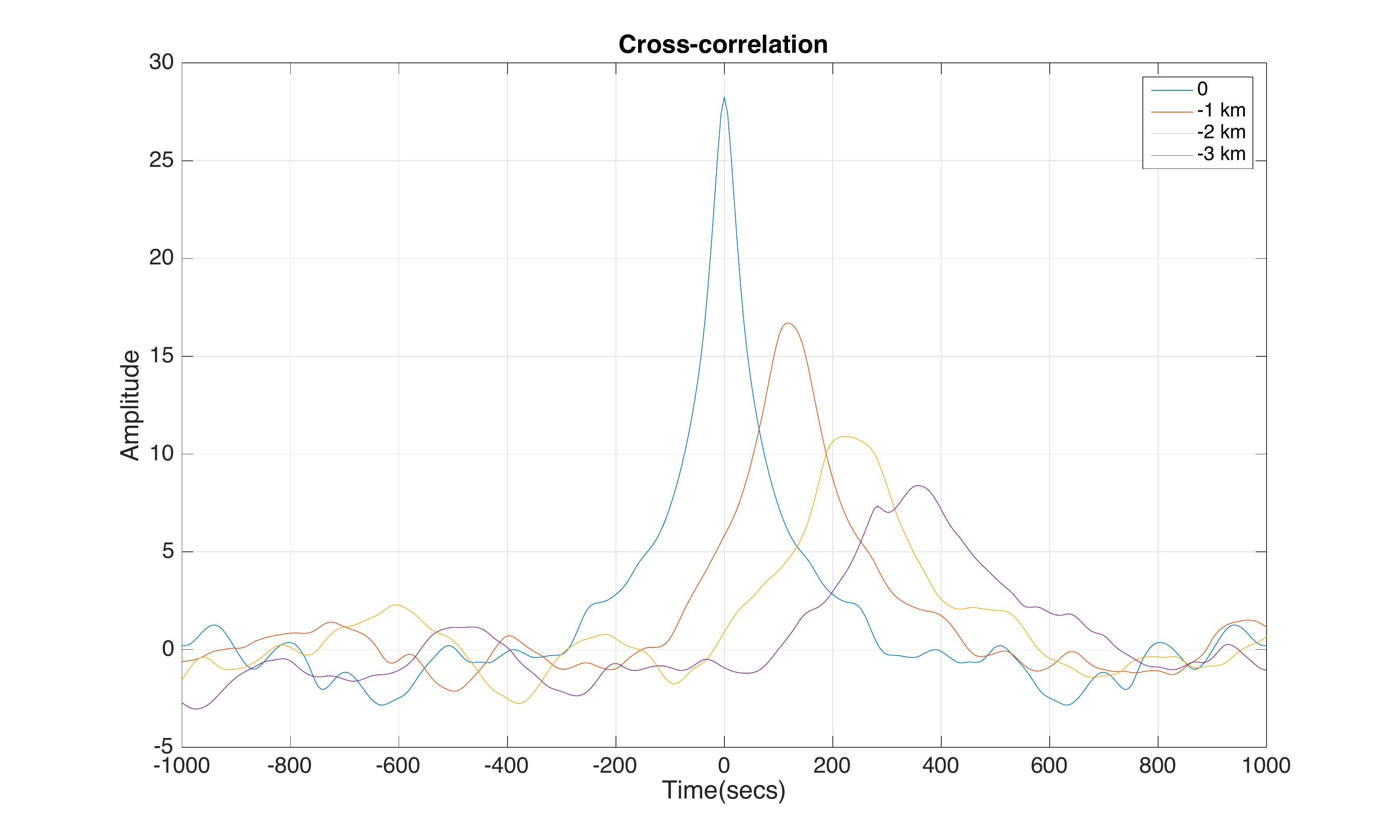}
\caption{(Top left) The cross correlation of the pressure observed in the center of the LES grid and the pressure observed at distances from the center of the grid (in the W-E direction). Each line corresponds to a different distance ranging from 0 m to +3000 m where 0 is at the center of the grid and positive distances are in the easterly direction. (Top right) A close-up of the same figure showing clearly the decreasing amplitude of the correlation and the increasingly negative time lag. (Bottom) A close-up of the cross correlation of the pressure observed in the center of the LES grid and the pressure observed at distance from the center of the grid (in the E-W direction). Each line corresponds to a different distance ranging from 0 m to -3000 m where 0 is at the center of the grid and negative distances are in the westerly direction. The decreasing amplitude of the correlation and the increasingly positive time lag can be seen.}
\label{fig:PressureCrossCorrelation}
\end{center}
\end{figure}

First, the variation of the correlation of the (detrended) pressure signals with distance is considered. In the following figures, the distance is increased in steps of 1 km in the W-E direction from 0 to 3 km, and the variation of the cross-correlation of the signals as a function of distance is investigated (top figures in \fig{PressureCrossCorrelation}).  The cross correlation (particularly the close-up) shows clearly the decreasing amplitude of the correlation with increasing distance. Additionally, the increasing time lag with distance is also evident. In this example, the time lag is increasingly negative as the distance increases in the W-E direction. This is due to the fact that the pressure disturbances pass the center of the grid first, and reach the easterly points at a later time (the larger the distance, the longer the time, and thus the larger the time lag). This is verified in the lower figure of \fig{PressureCrossCorrelation} in which the distance increases again in steps of 1 km, but this time in the E-W direction from 0 to -3 km. Here the time lag is positive, indicating that the disturbances reach the center of the grid at a later time than the more westerly points.

\begin{figure}[h!]
\begin{center}
\includegraphics[scale=0.15]{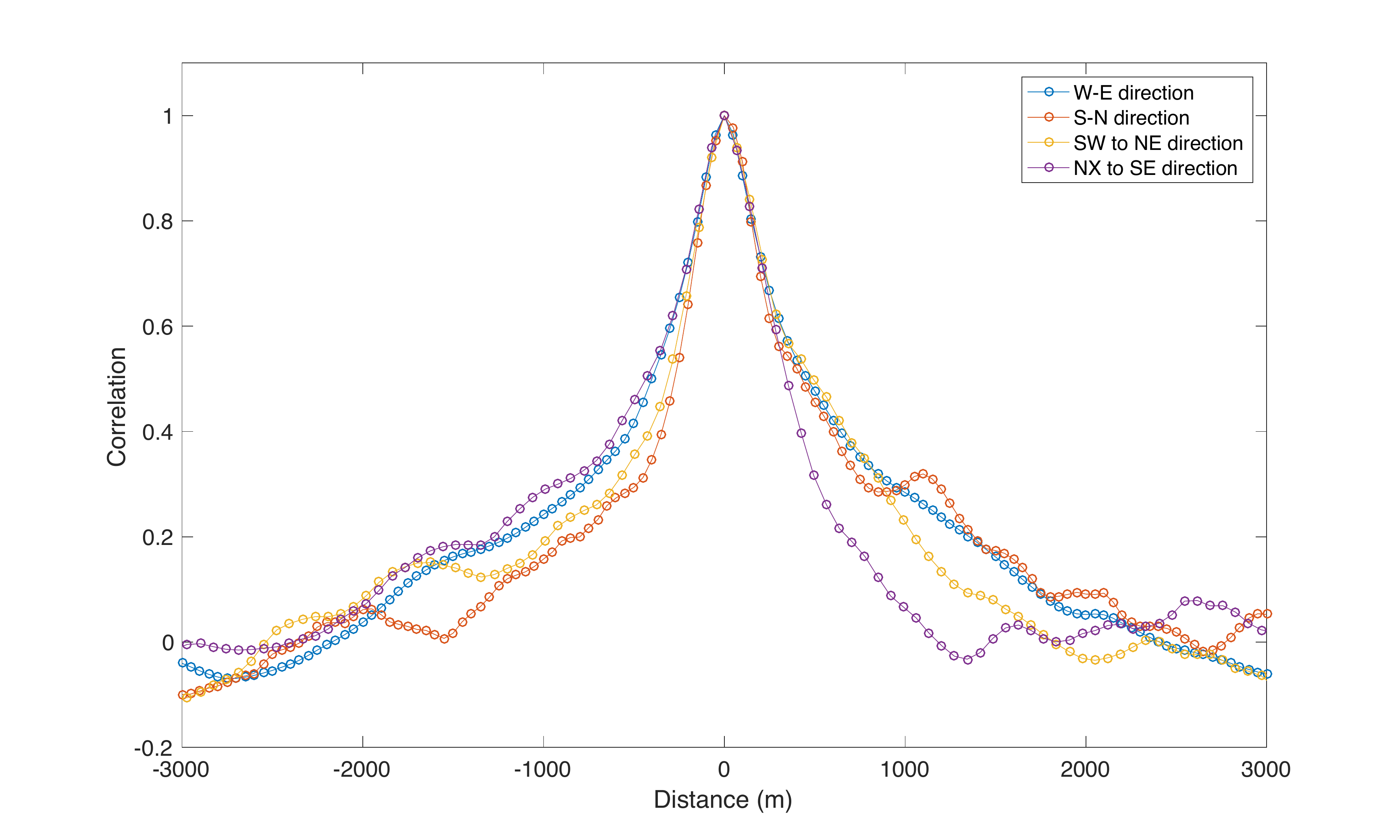}
\includegraphics[scale=0.157]{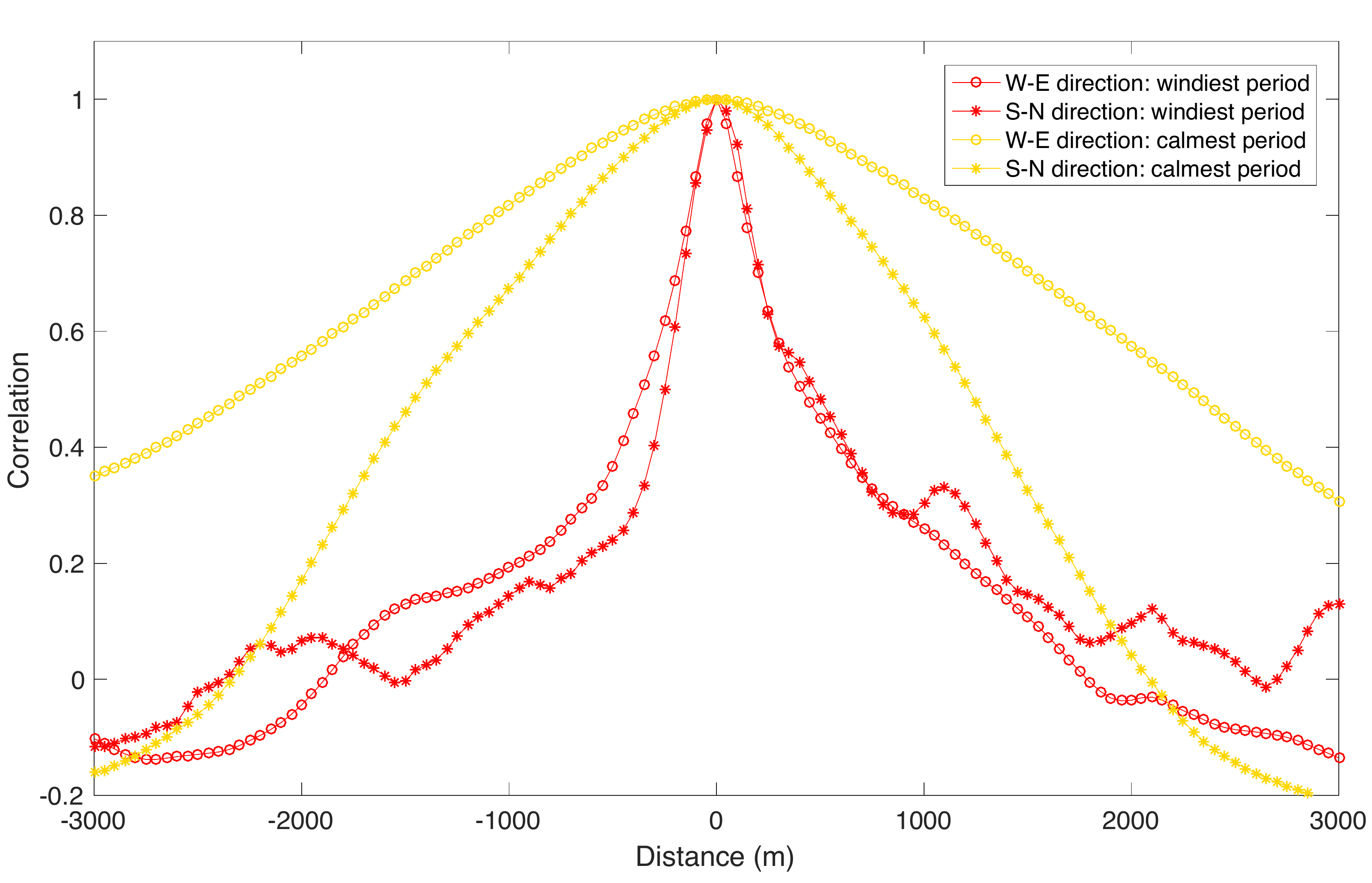}
\caption{Spatial correlation of the pressure field with itself in different directions. Correlation between the pressure signal at the center of the field at different distances in the (blue) W-E direction, (orange) the S-N direction, (yellow) the SW to NE direction, and (purple) the NW to SE direction. Note that the pressure field propagates in the W-E direction and the entire data set is used. (Right) The correlation as a function of distance for the windiest 2h period (red) and the calmest 2h period (yellow) in the W-E direction (circles) and the S-N direction (stars). Distance intervals of 50 m are used \ie the highest resolution possible.}
\label{fig:PressurePearson}
\end{center}
\end{figure}

Next, to determine more precisely how the amplitude of the correlation varies with distance, the Pearson Correlation Coefficient between the two pressure signals is used. This is a measure of the linear correlation between two variables, giving a value between +1 and -1 inclusive, where 1 is total positive correlation, 0 is no correlation, and -1 is total negative correlation.  For each distance from the center of the grid, the degree of correlation is calculated using this correlation coefficient and the degree of correlation is then plotted as a function of distance, considering intervals of 50 m \ie the highest resolution possible. The results are shown in \fig{PressurePearson}.

One could possibly expect the correlation distance to be linked to the wind speed. To test this hypothesis, two periods of data are considered: the windiest (most turbulent) two-hour period in the dataset (from 12.9h to 14.9h) and the calmest (least turbulent) two-hour period on the dataset (from 18.2h to 20.2h).  These are shown in \fig{WindPressurecenterOfGrid}. Note that there may also be variations of regional winds in addition to the turbulence, but these are not considered here, as the mean background wind is constant. 

\Fig{PressurePearson} shows the correlation as a function of distance in both the W-E and S-N directions for the windy and quiet periods.  It can immediately be seen that the pressure correlation is much smoother and more regular during the quiet periods than during the windy periods. This implies that, in calm conditions, the single-pressure measurement is more representative of the large-scale pressure field, particularly in the wind direction. However, when windy, small-scale turbulence results in a reduced correlation between the pressure signals, and the single-pressure measurement becomes less representative of the pressure field. Unfortunately, however, given that the pressure noise increases with the wind speed, the windy period is also the time when the seismic pressure noise will be highest  (\fig{WindyCalmPSD}).

These correlation results already shed light on the perspectives of pressure decorrelation. Sorrell's theory (described in detail in \sect{sorrells}), assumes that a pressure field is carried by the wind, and that the typical wavelength generating the tilt is provided by $c$ $\times$ $\tau$, where $c$ is the ambient wind speed and $\tau$ is the period. For 10 m/s wind speed and a 50-second period, the wavelength is about 0.5 km, and corresponds to distances over which a significant correlation remains, as noted by \fig{PressureCrossCorrelation}.  These correlations leave us, therefore, with good prospects on the pressure decorrelation, as will be demonstrated later.

\section{Simulating the seismic signal from pressure variations}
\label{s:seismic}

\subsection{Green's function approach}

\cite{lorenz2015} have shown that the shape and amplitude of dust devil seismic signals can be modeled with a simple quasi-static point-load model of the negative pressure field associated with the vortices acting on the ground as an elastic half-space. This point-load ground deformation approach has been validated via comparison with in-situ seismic and pressure measurements of terrestrial dust devils. The same approach as \cite{lorenz2015} is, therefore, used here for the ground deformation calculations. The ground is modeled as an elastic half-space with properties of a Martian regolith (\tbl{regolith}). This model gives a worst-case estimate for the seismic noise as the presence at some (shallow) depth of a harder layer would significantly reduce quasi-static displacement and tilt effects. The following Boussinesq-Cerrutti solution is then used to calculate the displacement of the ground at the SEIS feet. 

Assume that we have a point force ${ F}=F_1 { e_1}+ F_2 { e_2}+F_3 { e_3}$ that is applied at the point ${\boldsymbol \xi} = \xi_1 { e_1}+ \xi_2 { e_2}+ \xi_3 { e_3}$ and ${ \Lambda}= \Lambda_1 { e_1}+ \Lambda_2 { e_2}+ \Lambda_3 { e_3}$  is some arbitrary point in the half-space $\Lambda_3 \ge 0$. Green's tensor for displacements ($G_{ik}$), defined by the relation $u_i= \sum_{k} G_{ik}  F_k$, may be written in Cartesian coordinates as \citep[solution from][]{landau1970}:

\begin{center}
\vspace{0.2 cm}
$G_{ik} = \frac{1}{4\pi \mu}$   
\begin{footnotesize}
$\left[\begin{array}{cccc}
\frac{b}{r} + \frac{x^2}{r^3} - \frac{ax^2}{r(r+z)^2} - \frac{az}{r(r+z)} & \frac{xy}{r^3} - \frac{ayx}{r(r+z)^2}  &  \frac{xz}{r^3} - \frac{ax}{r(r+z)}        \\
\frac{yx}{r^3} - \frac{ayx}{r(r+z)^2} & \frac{b}{r} + \frac{y^2}{r^3} - \frac{ay^2}{r(r+z)^2} - \frac{az}{r(r+z)}  &  \frac{yz}{r^3} - \frac{ay}{r(r+z)}         \\
\frac{zx}{r^3} - \frac{ax}{r(r+z)} & \frac{zy}{r^3} - \frac{ay}{r(r+z)}  & \frac{b}{r} + \frac{z^2}{r^3}
\end{array}\right]$
\end{footnotesize}
\vspace{0.2 cm}
\end{center}

\noindent where $x = \Lambda_1 - \xi_1$, $y = \Lambda_2 - \xi_2$, $z = \Lambda_3 - \xi_3$, and $r$ is the magnitude of the vector between ${ \Lambda}$ and ${\boldsymbol \xi}$, $a = (1-2\nu)$ and $b = 2(1-\nu)$, $\nu$ is Poisson's ratio and $\mu$ is the shear modulus.  For our calculations, this derivation can be simplified as both the point force and the arbitrary point of measure are on the surface \ie $z$ = 0.

\begin{figure}[h!]
\begin{center}
\includegraphics[scale=0.5]{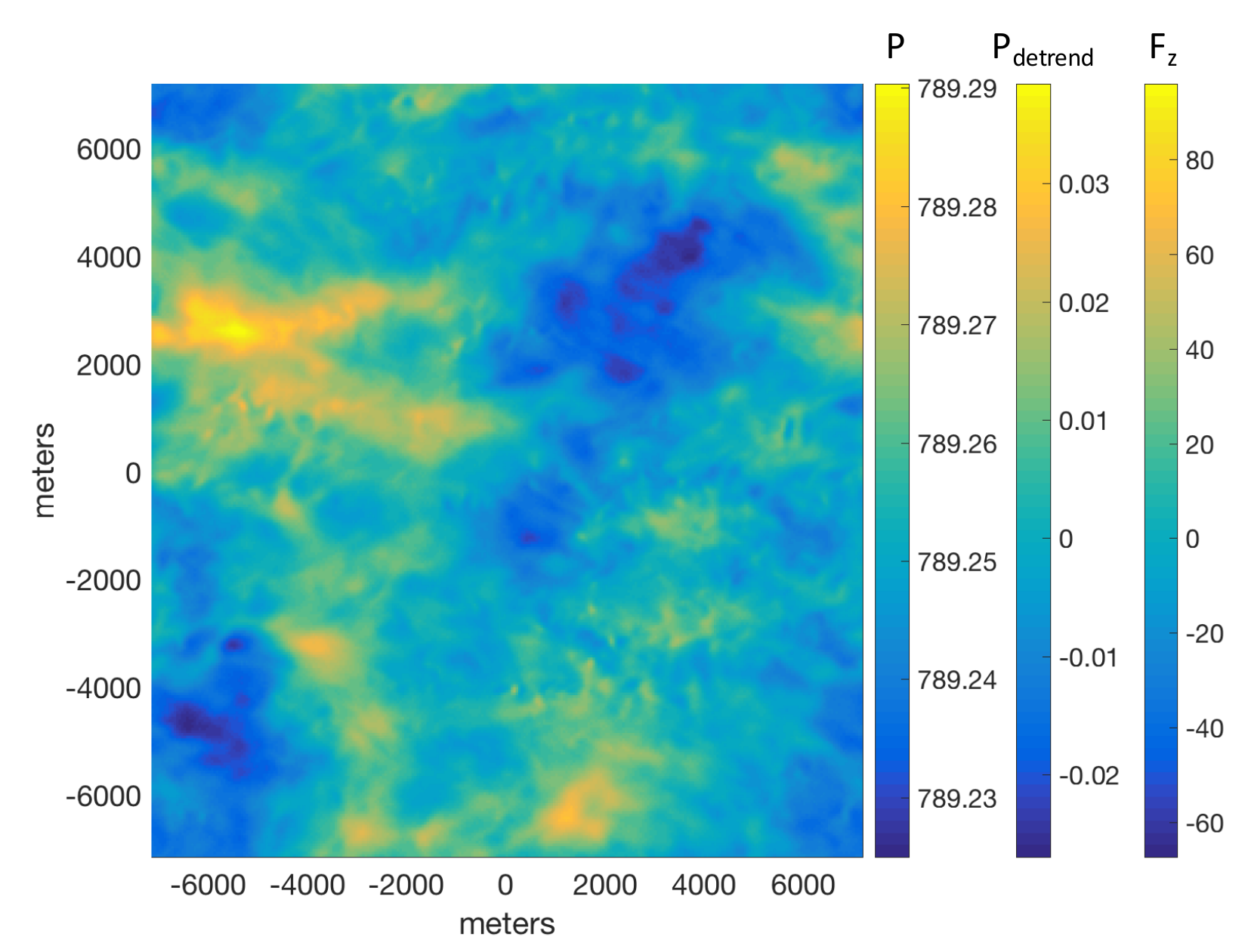}
\caption{Figure showing the pressure ($P$, in Pa), detrended pressure ($P_{detrend}$, in Pa) and vertical force ($F_z$, in N) variations across the LES grid at one instant in time.  In these figures, North is aligned with the $y$-axis and East is aligned with the $x$-axis.}
\label{fig:PressureForceExample}
\end{center}
\end{figure}

For every section of the LES grid, the variation of the vertical force exerted on the ground at the center of the section of the grid can be given by the detrended value of the pressure of the grid section times the surface area of the grid section (\fig{PressureForceExample}). Then, the displacement of the ground at the seismometer feet will be a sum of the displacements caused by each section of the grid (each considered to be a point source in Green's function approximation). { We do not correct for the free-air anomaly as this is expected to become significant only at periods longer than those considered here \citep[$>$ 1000 s;][]{kenda2016}.}


\begin{figure}[h!]
\begin{center}
\includegraphics[scale=0.25]{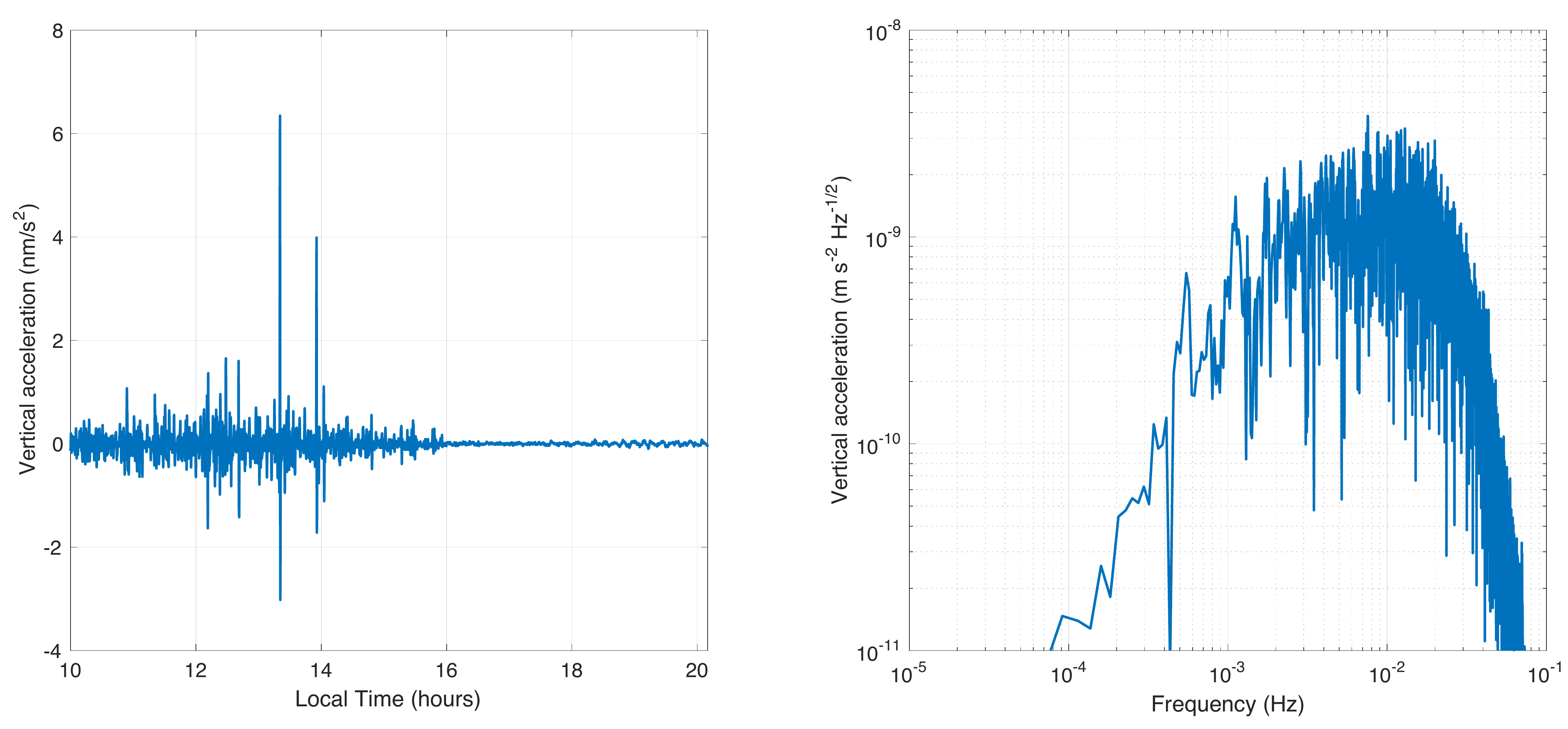}
\caption{(Left) The direct vertical acceleration of the ground induced by the pressure fluctuations as a function of time. (Right) The amplitude spectral density of the vertical ground acceleration.}
\label{fig:VerticalAcc}
\end{center}
\end{figure}

\begin{figure}[h!]
\begin{center}
\includegraphics[scale=0.15]{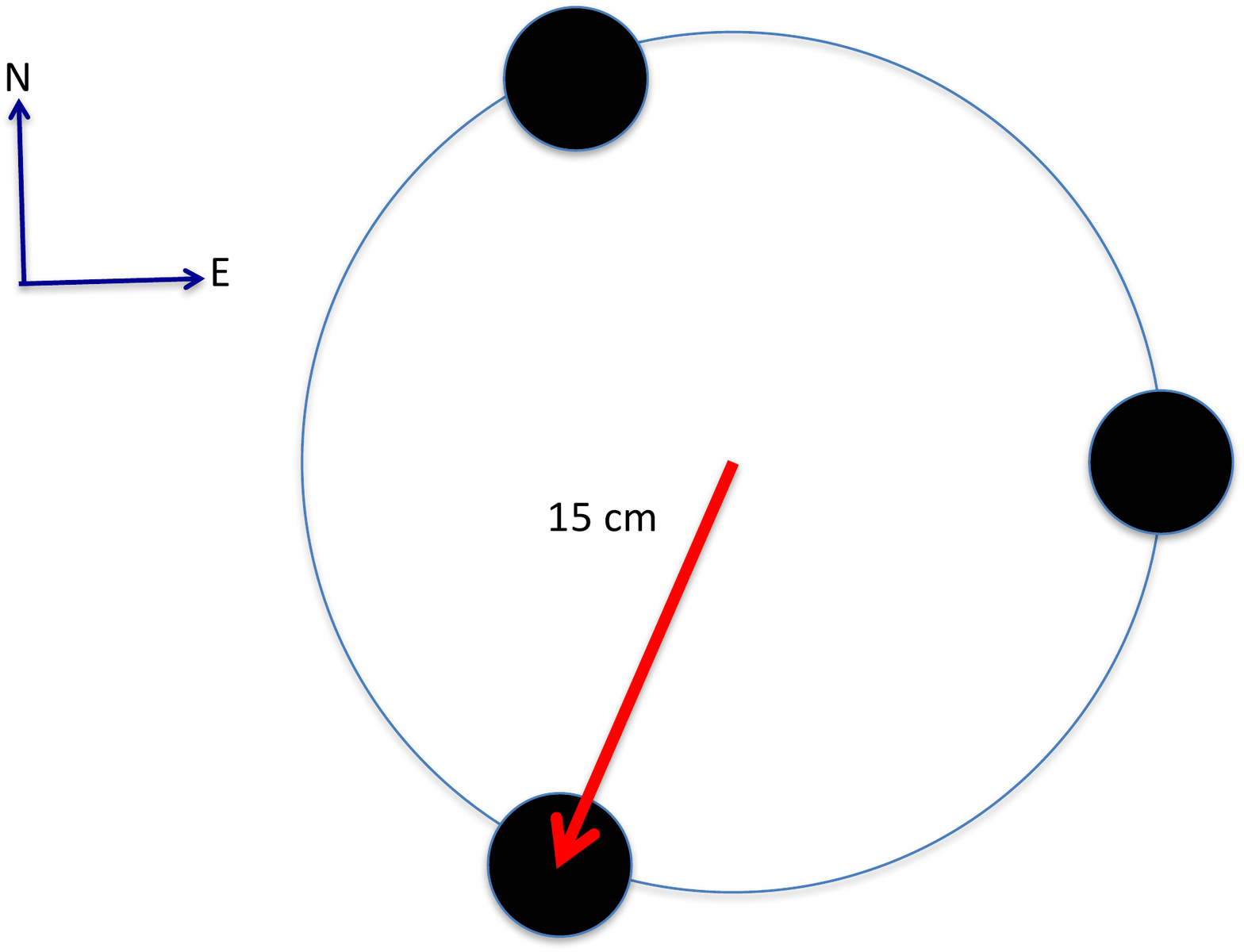}
\includegraphics[scale=0.3]{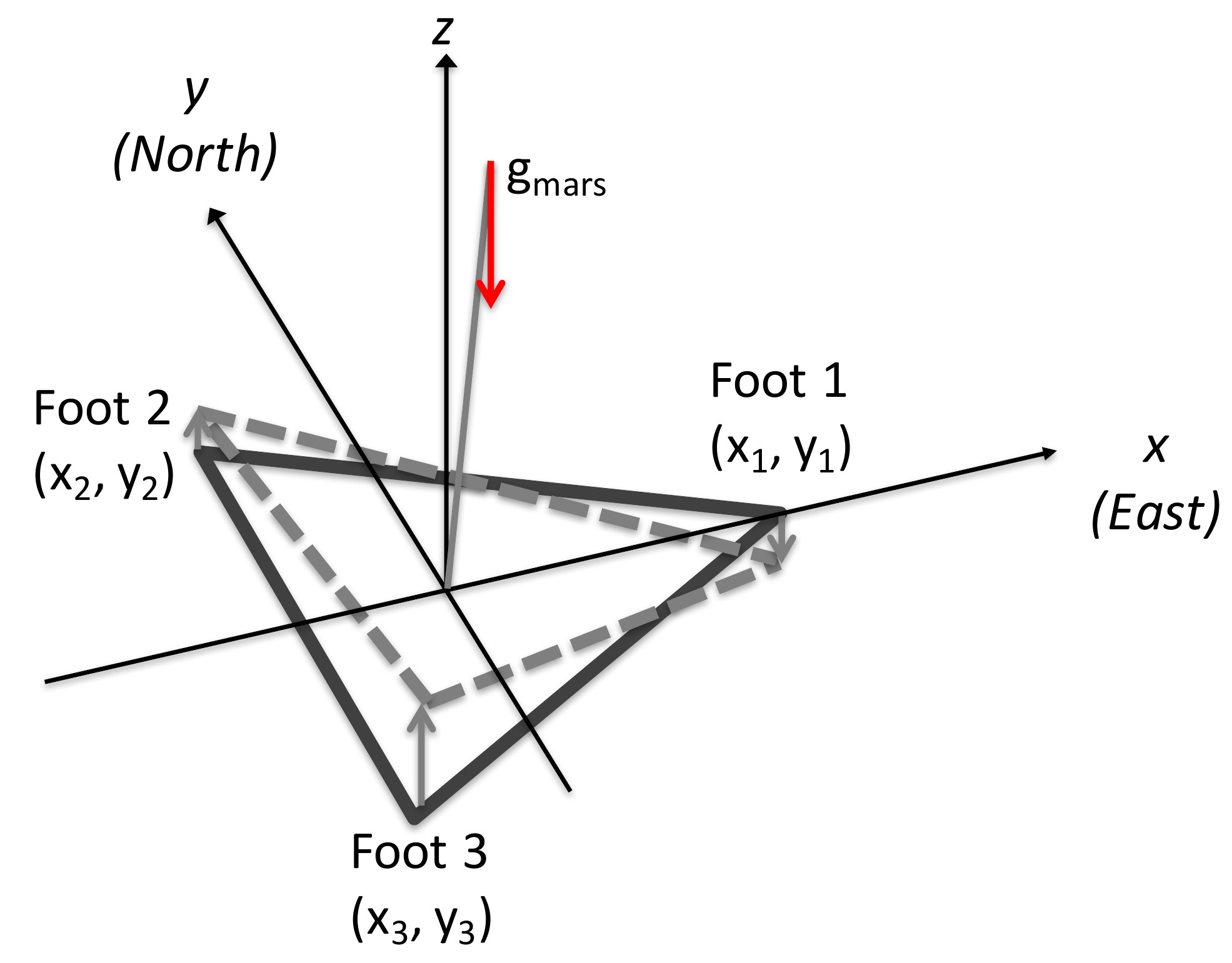}
\caption{(Left) Schematic of SEIS feet location and orientation. (Right) Schematic explaining the tilt noise seen by the seismometer due to the different vertical displacements of the three feet in the gravity field.}
\label{fig:FeetLocation}
\end{center}
\end{figure}

\begin{figure}[h!]
\begin{center}
\includegraphics[scale=0.3]{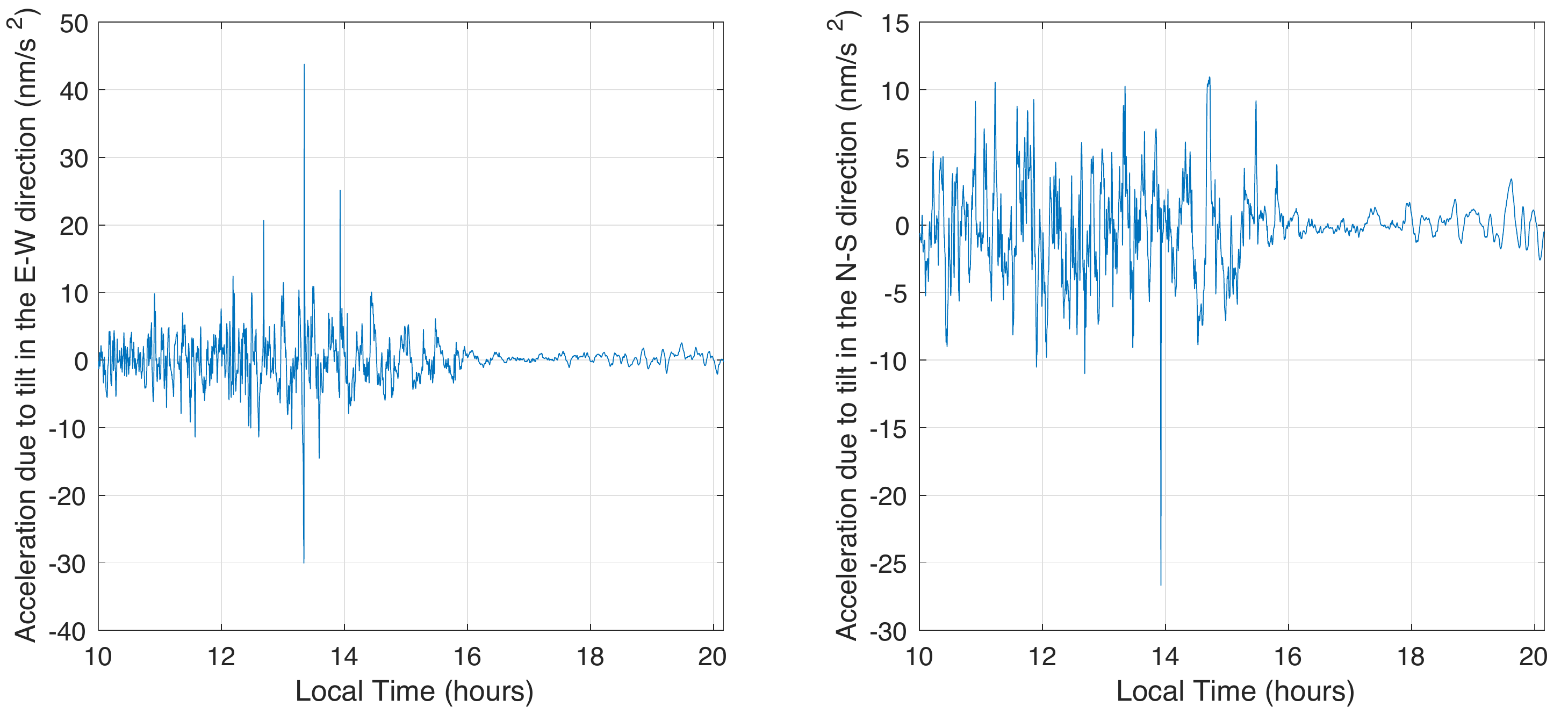}
\includegraphics[scale=0.2]{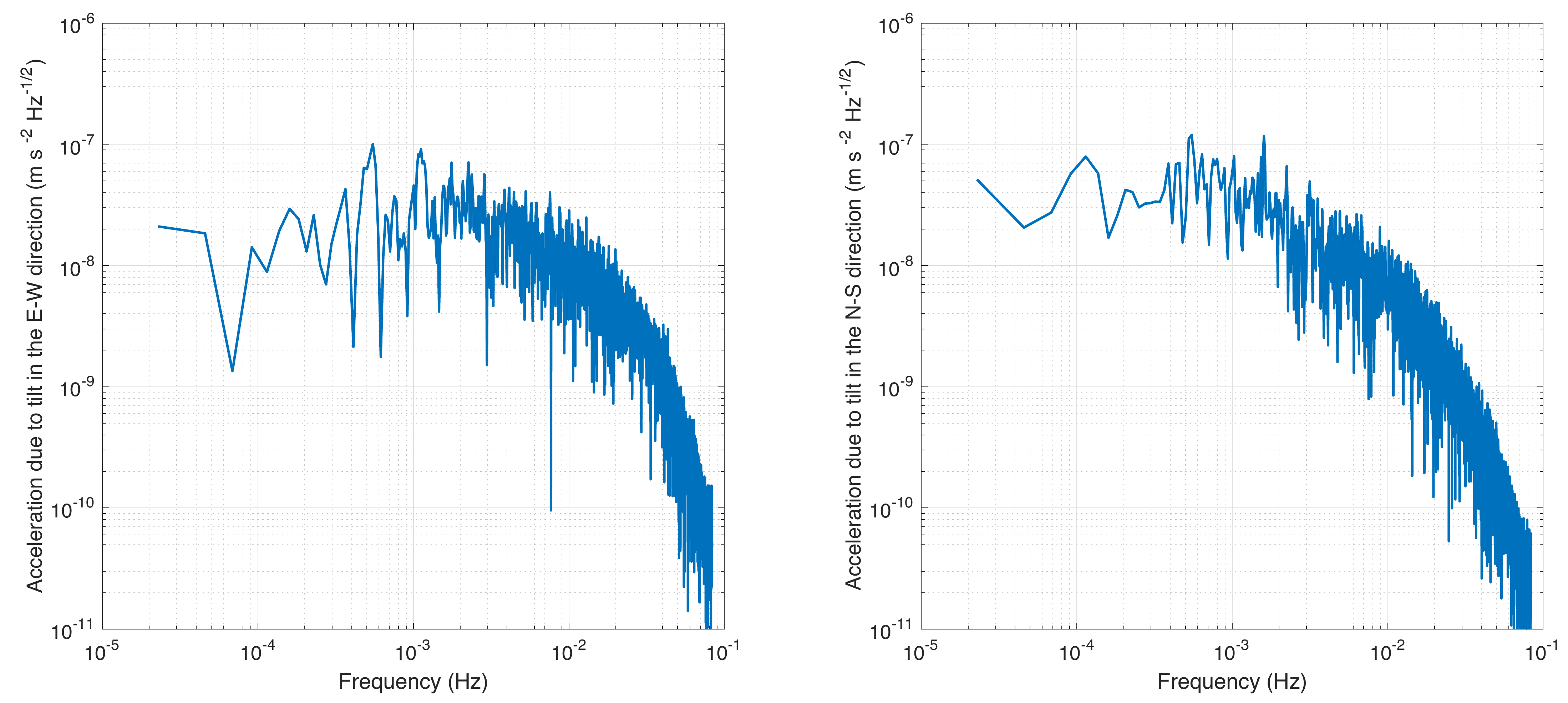}
\caption{The tilt acceleration in the E-W direction (upper left) and in the N-S direction (upper right) as a function of time due to the pressure fluctuations predicted by the LES (+N, +E). The amplitude spectral density of the tilt acceleration in the (lower left) E-W direction and (lower right) N-S direction. The time period is from 10am local time (just after stabilization of the simulations) to just after 8pm local time.}
\label{fig:TiltAcc}
\end{center}
\end{figure}

The vertical acceleration of the ground induced by the pressure fluctuations is shown in \fig{VerticalAcc}. The tilt of SEIS can be calculated, taking into account the different vertical displacement of the ground under the three SEIS feet (\fig{FeetLocation}). Assuming that the tilt is small, the magnitude of the acceleration due to the tilt in the E-W (+E) and N-S (+N) directions ($A_{EW}$ and $A_{SN}$, respectively) can be approximated by:

\begin{align}
A_{EW} & = g_{mars} \frac{(\Delta z_2 +  \Delta z_3)/2 - \Delta z_1}{ |x_2 - x_1|}\\
A_{NS} & = g_{mars} \frac{\Delta z_3 -  \Delta z_2}{ |y_3- y_2 |}\\
\end{align}

\noindent where $\Delta z_1$, $\Delta z_2$  and $\Delta z_3$ are the vertical displacements of the ground under SEIS feet 1, 2 and 3, respectively, $x_1$, $x_2$ are the $x$ coordinates of the feet 1 and 2, $y_2$, $y_3$ are the $y$ coordinates of the feet 2 and 3.  The center of SEIS is assumed to be perfectly centered in the LES field, and the SEIS feet are located at a radius of 15 cm from the geometric center of SEIS as in the following schematic (\fig{FeetLocation}). The tilt acceleration observed on SEIS as a result of the ground tilt in the E-W and N-S directions due to the LES pressure field is calculated (\fig{TiltAcc}). 

\begin{figure}[h!]
\begin{center}
\includegraphics[scale=0.3]{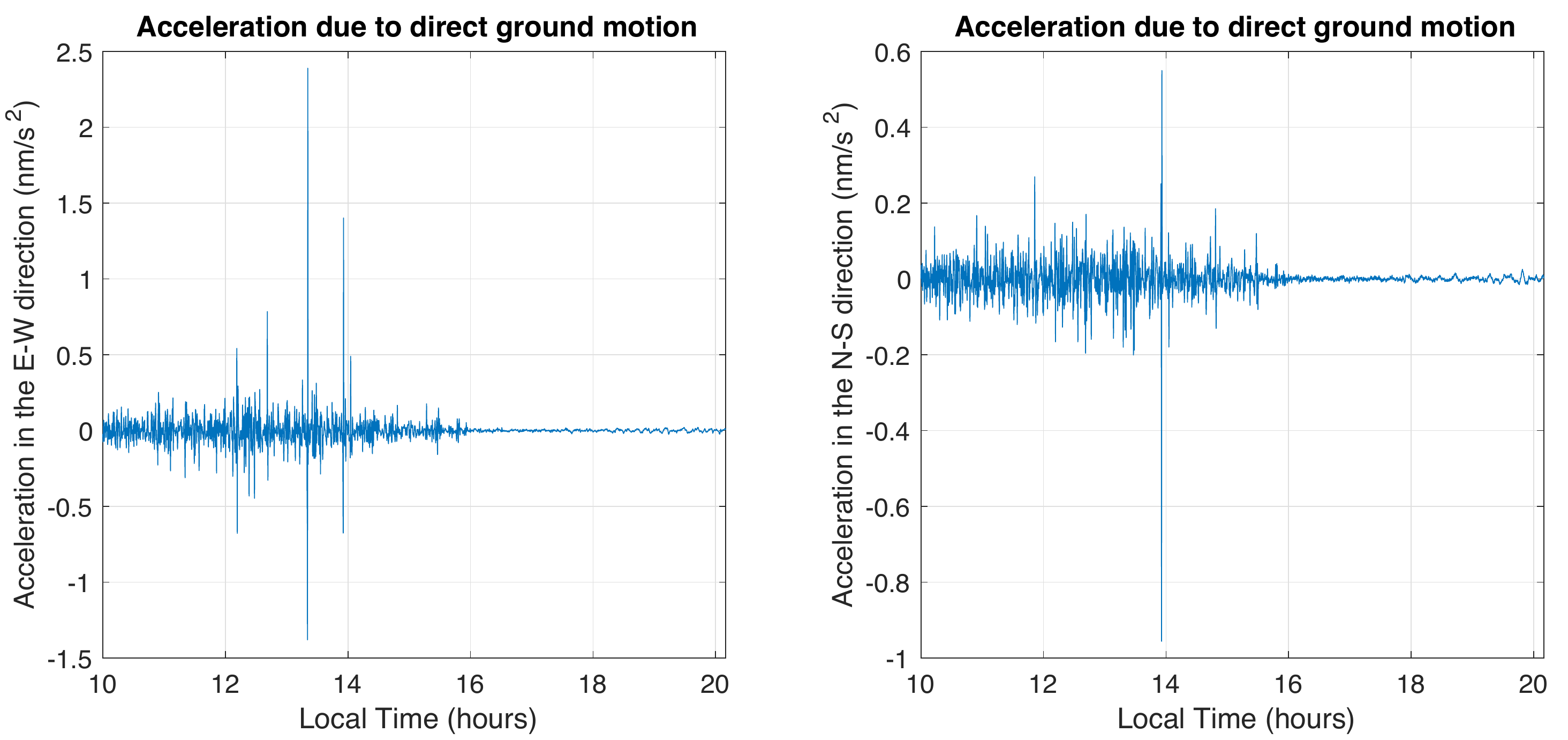}
\caption{The direct horizontal acceleration of the ground induced by the pressure fluctuations in the E-W direction (left) and in the N-S direction (right).}
\label{fig:directacc}
\end{center}
\end{figure}

The real horizontal acceleration of the ground induced by the pressure fluctuations is also calculated, in addition to the ground tilt only from the vertical displacement (\fig{directacc}). This direct acceleration contribution has a mean magnitude of 0.03 nm/s$^2$ and 0.02 nm/s$^2$ in the E-W and N-S directions, respectively. This is two orders of magnitude smaller than the contribution to the acceleration from the ground tilt and is thus negligible. Therefore, only the ground tilt is used for the horizontal accelerations in the subsequent analyses. 



\begin{figure}[h!]
\begin{center}
\includegraphics[scale=0.2]{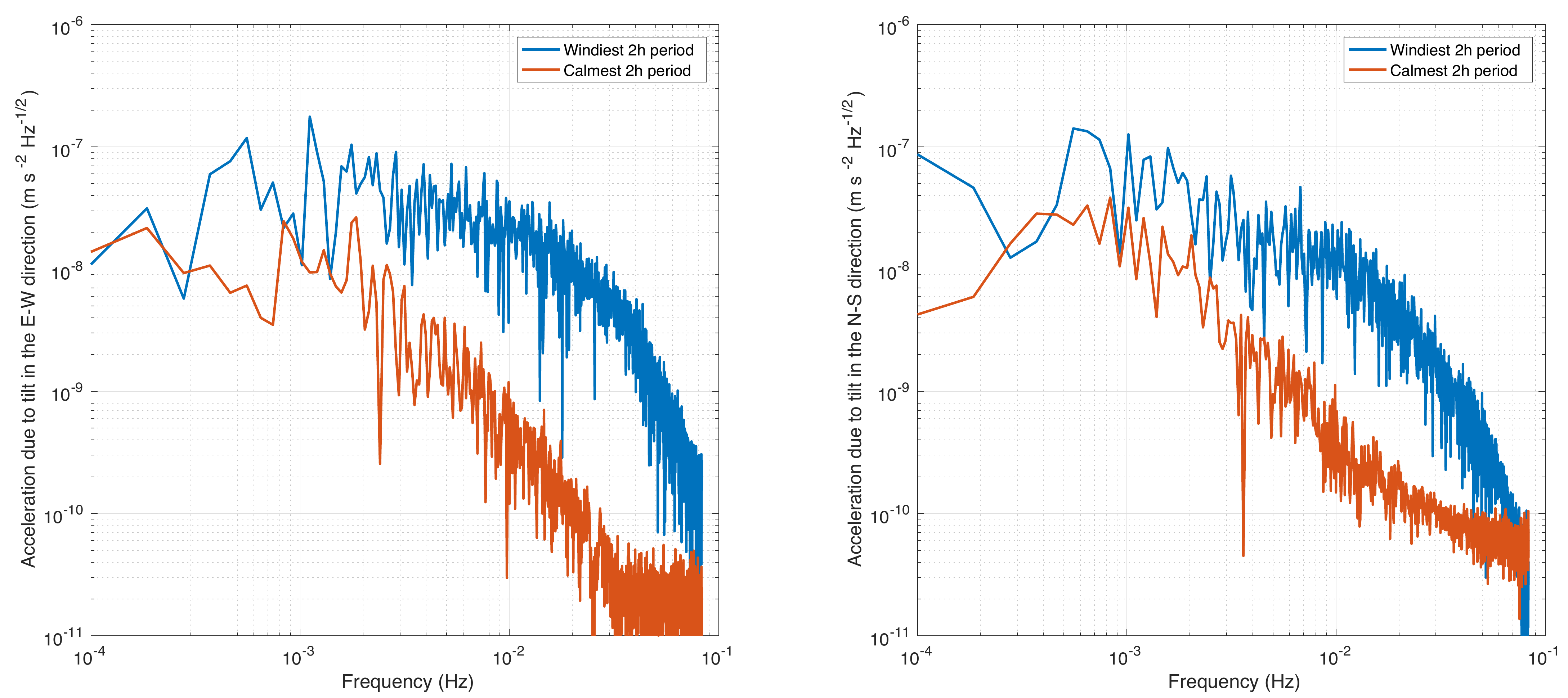}
\caption{The amplitude spectral density of the acceleration due to tilt in the (a) E-W direction and the (b) N-S direction during the windiest 2h period (blue) and the calmest 2h period (orange).}
\label{fig:WindyCalmPSD}
\end{center}
\end{figure}

\Fig{WindyCalmPSD} shows how the amplitude of the tilt noise varies in periods of time with more or less turbulence. The windiest (most turbulent) period generates significantly larger vertical accelerations than during the calm period.

\newpage
\subsection{Comparison with a spectral approach}

A different computation of the tilt induced by pressure fluctuations is detailed in \cite{kenda2016}, who derived the displacement at the surface from the same LES pressure field and the response of the ground to static loading for various subsurface models. The 3D quasi-static displacement $\bar{u}$ is obtained by convolution of the source, \ie the pressure fluctuation $P$, with the response function $\bar{R}$ as follows:

\begin{equation} 
\bar{u}(x,y,t)=\int e^{i(\omega t - k_x x-k_y y)}\bar{R}(k_x,k_y) P(k_x,k_y,\omega) \mbox{d}\omega \mbox{d} k_x \mbox{d} k_y .
\end{equation} 

\noindent Here $x$ and $y$ are the two horizontal cartesian coordinates, $t$ is time, and $k_x$, $k_y$ and $\omega$ are the corresponding wave numbers and angular frequency. The resulting acceleration fields are then corrected for the free-air anomaly and the tilt of the ground to obtain the seismometer acceleration at each grid-point.

\begin{figure}[h!]
\begin{center}
\includegraphics[scale=0.21]{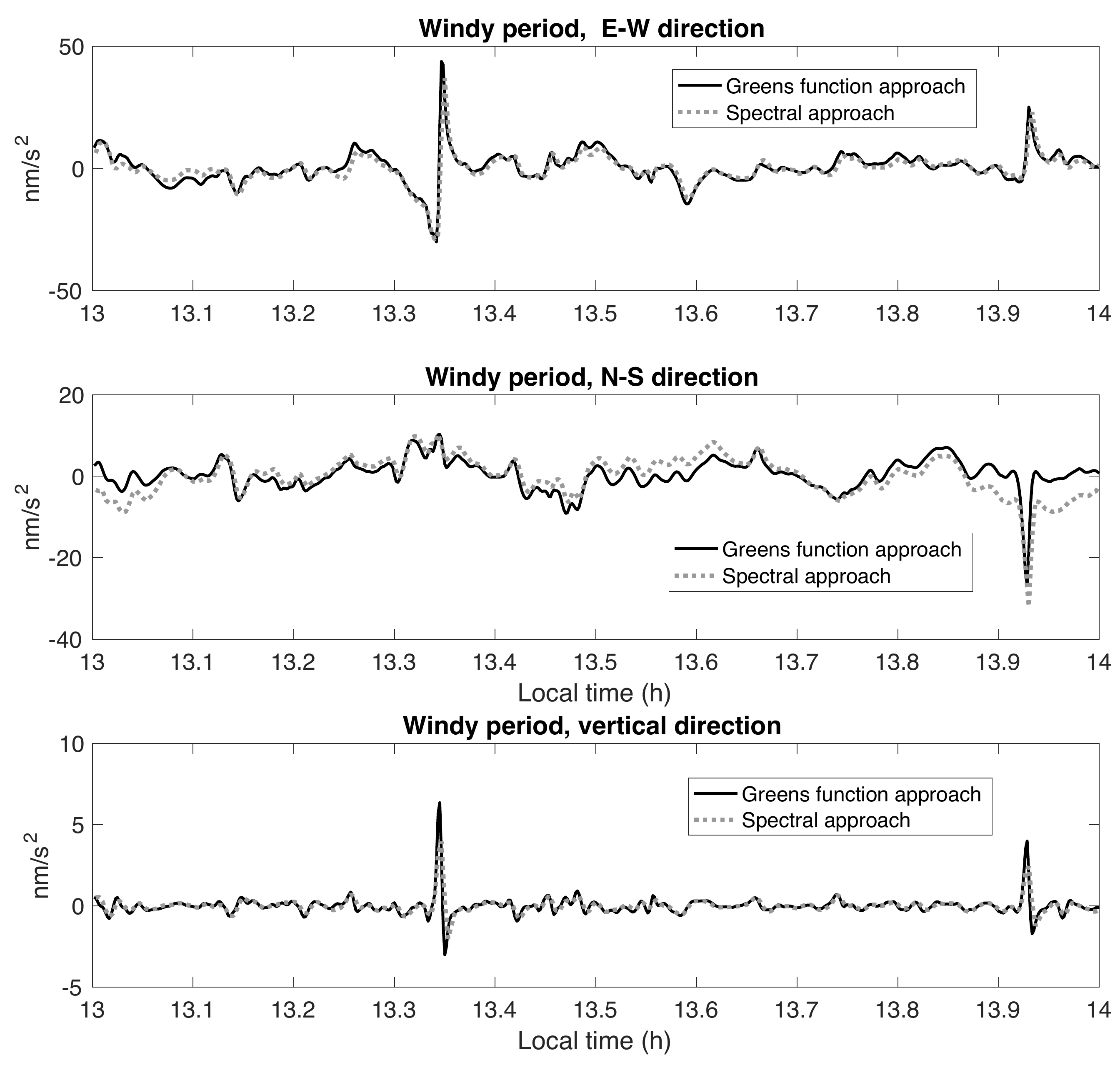}
\includegraphics[scale=0.21]{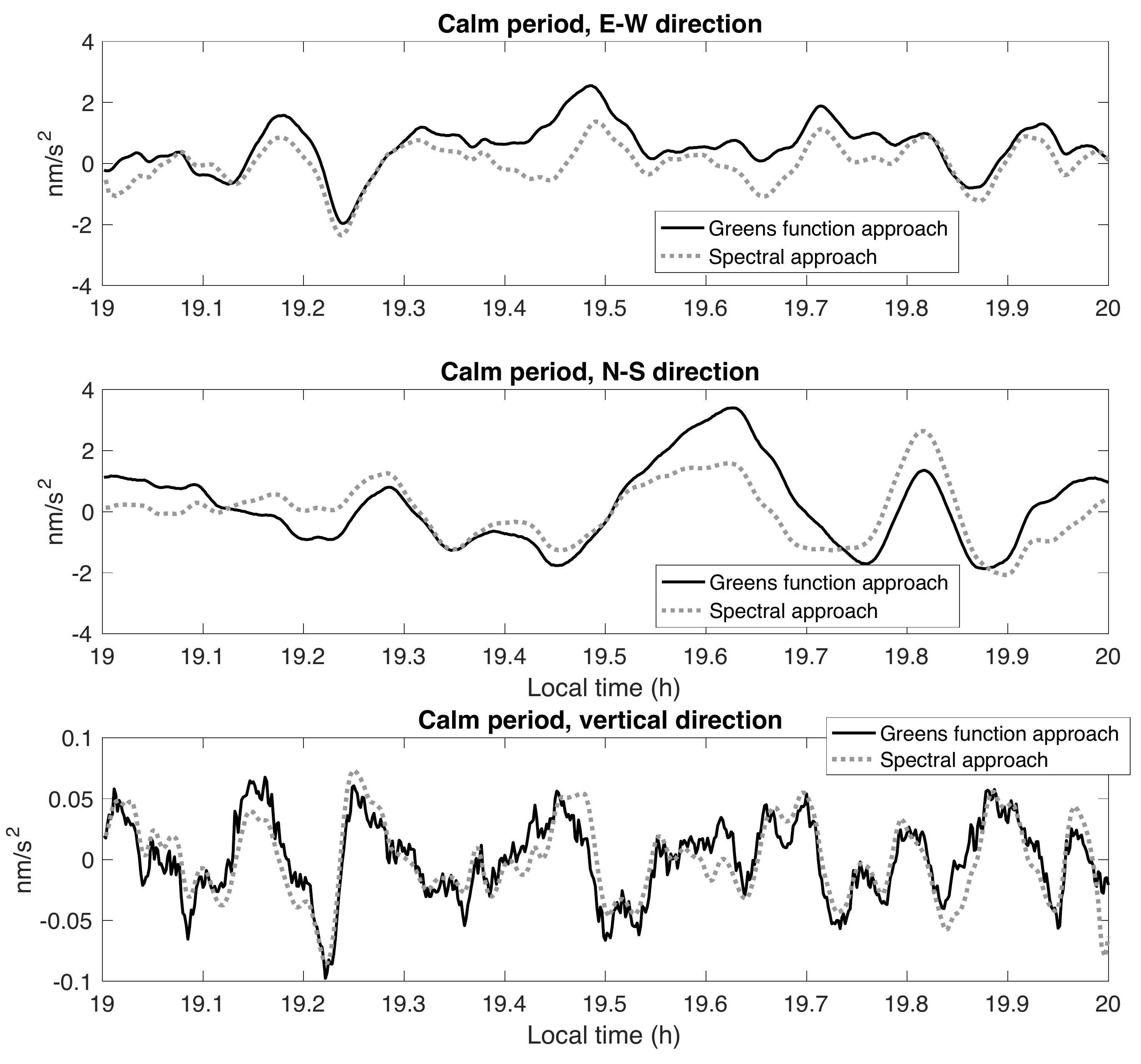}
\caption{Comparison of the E-W, N-S and vertical accelerations calculated using Green's function method (black) and the spectral approach (grey dotted) for 1h during the windiest period (left) and 1hr during the calmest period (right).  }
\label{fig:ModelComparison}
\end{center}
\end{figure}

\Fig{ModelComparison} provides the comparison of the E-W, N-S and vertical accelerations calculated using Green's function method and this spectral method for one hour during the windiest period and one hour during the calmest period (right).  The two models give very similar results for all time periods, and the differences between the two methods are small, on the order of nm/$\mbox{s}^2$. { These small differences may be due to the different methods used to detrend the pressure field before performing the ground displacement calculations.  It can also be seen that Green's function approach generates higher frequencies that are not observed in the spectral approach; this is particularly evident for the vertical velocity (lower right plot in \fig{ModelComparison}). These high frequencies are likely a result of the discretization of the pressure field in Green's function approach. However, despite small differences, the good agreement between these two independently developed methods cross-validates the two approaches and ensures their robustness. }

\section{Comparison with Sorrells' method}
 \label{s:sorrells}

An alternative method to compute the seismic noise induced by atmospheric pressure fluctuations is based on Sorrells' theory \citep{sorrells1971,sorrells1971b}. This method is particularly useful whenever single-point meteorological and seismic data are available, which will be the case for the InSight mission. Here this approach is described briefly and the results are compared to the outputs of Green's function method described above.

\cite{sorrells1971} computed the quasi-static ground displacement generated by pressure loading in the hypothesis that the pressure fluctuations are plane waves propagating at the ambient wind speed $c$, that is $k=\frac{\omega}{c}$, where $k$ is the wave number and $\omega$ the angular frequency. This formulation becomes especially simple in the case of a homogeneous half-space, which is assumed throughout this work. Indeed, for a half-space with density $\rho$ and seismic velocities $v_P$ and $v_S$, the vertical ground velocity $V$ and the surface tilt $T$ are proportional to the pressure fluctuation $P$ \citep{sorrells1971}:

\begin{equation}
V=-i\frac{c}{\rho}\frac{v_P^2}{2v_S^2(v_P^2-v_S^2)}P, \mbox{ } \mbox{ } T=g\frac{V}{c},
\label{e:sorrell}
\end{equation}

\noindent where $g$ is surface gravity. The real horizontal acceleration of the ground is also described by this model, but its numerical values turn out to be 10-100 times smaller than the tilt effect and may thus be neglected (as in Green's function approach, \fig{directacc}).

\begin{figure}[h!]
\begin{center}
\includegraphics[scale=0.2]{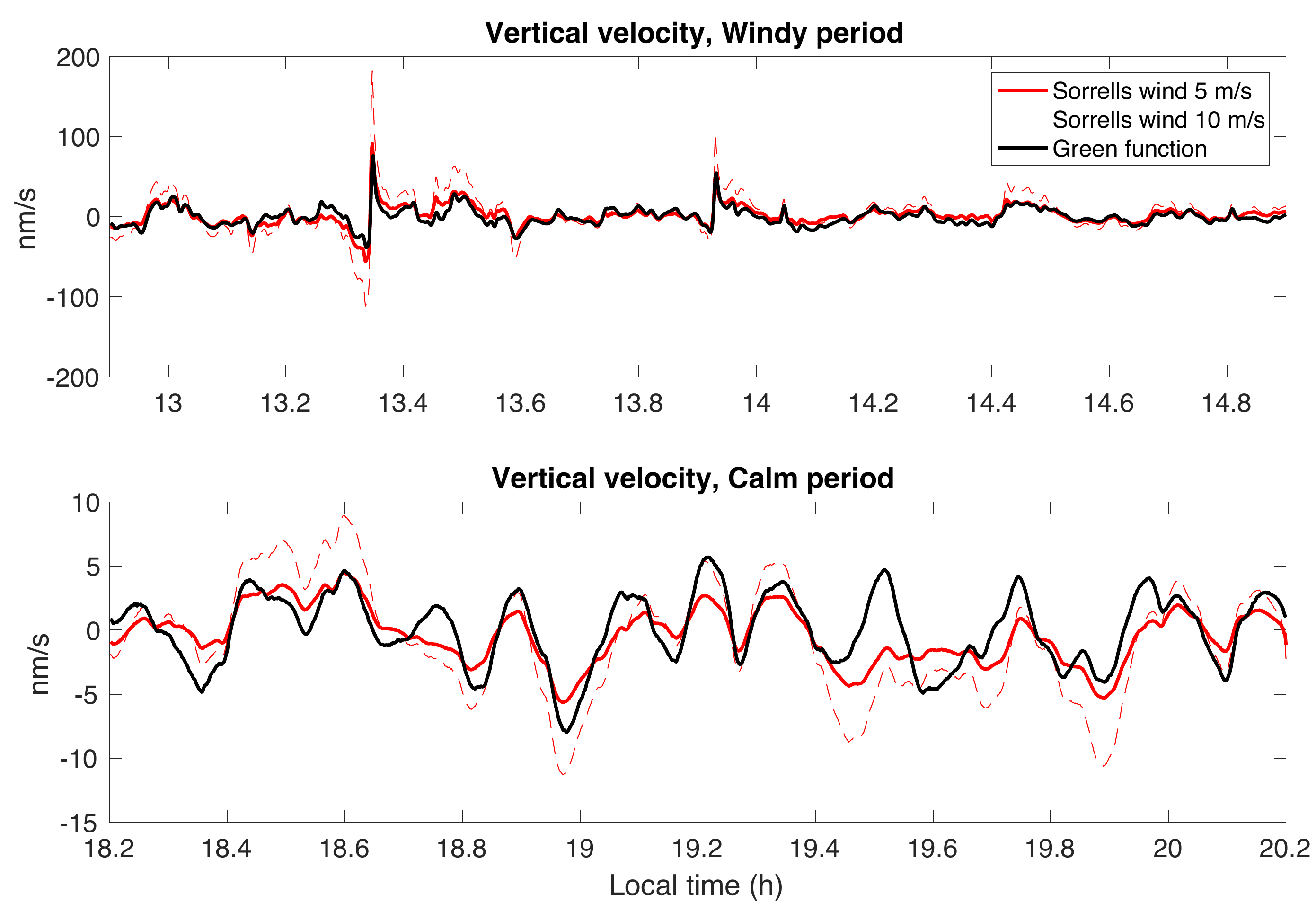}
\includegraphics[scale=0.2]{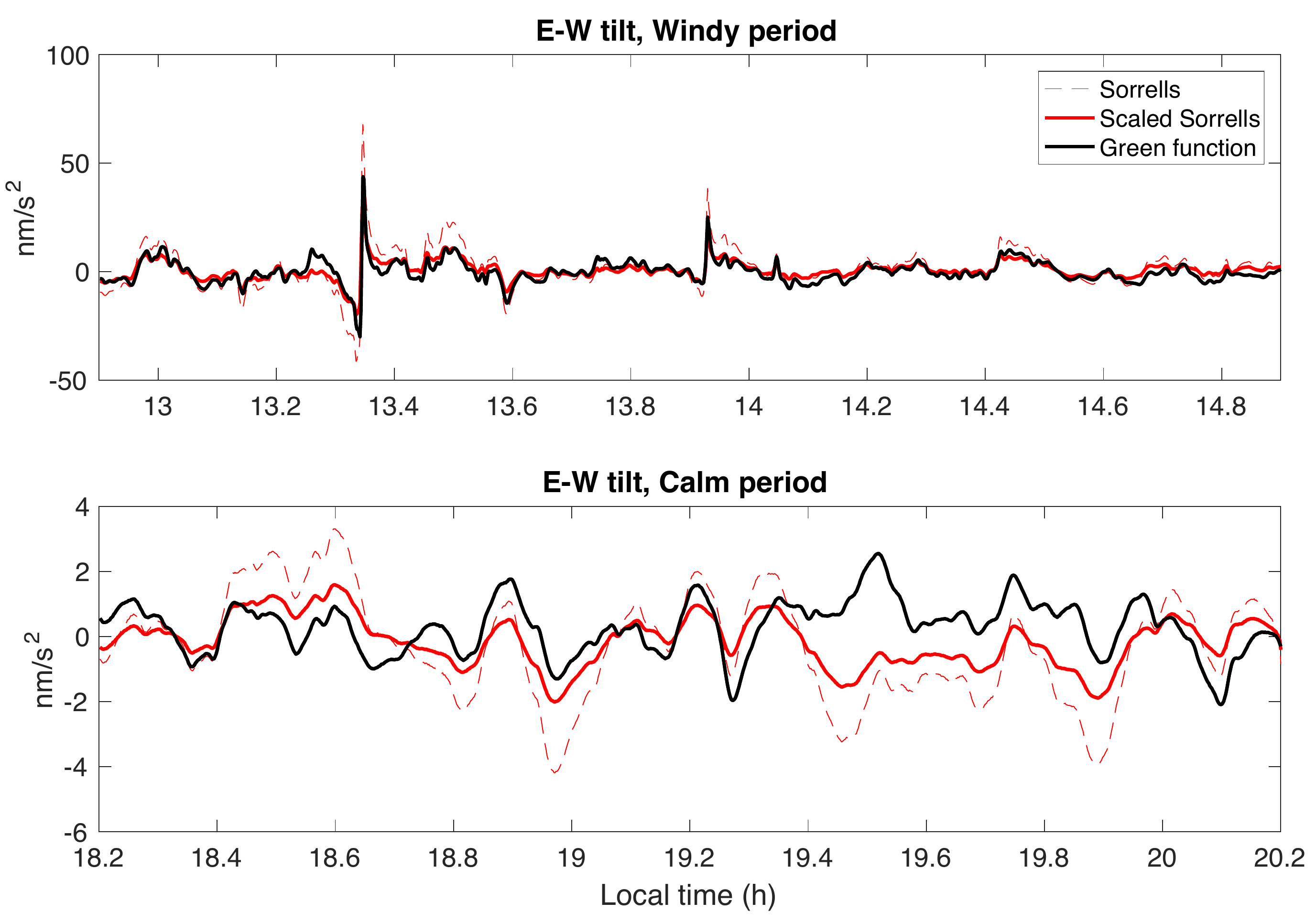}
\caption{(Left) Comparison of the (left) vertical ground velocities during the windiest and calmest periods calculated using Green's function method (black) and the single-station Sorrells approach with a wind speed of 5 m/s (solid red), and with a wind speed of 10 m/s (dashed red). (Right) Comparison of the ground tilt in the E-W direction during the windiest and calmest periods calculated using Green's function method (black) and the single-station Sorrells' approach (dashed red). 
The horizontal components of the Sorrells' seismograms are independent of background wind speed, but a scaling factor is needed to obtain a best fit to the data.  The solid red curve shows Sorrells' approach ground tilt results scaled to have the same peak amplitudes as Green's function results.}
\label{fig:ModelComparisonSorrells}
\end{center}
\end{figure}

Sorrells' approach is applied to the pressure time-series resulting from LES at the center of the grid, and the results are compared to the vertical velocity and the ground tilt computed with Green's function method (\fig{ModelComparisonSorrells}). The vertical velocities obtained with the two techniques compare very well for a wind speed of 5 m/s, the value measured at a height of 1.5 m from the surface, whereas the amplitude differs for a wind speed of 10 m/s, the value of the background wind imposed in the LES simulation. Concerning the tilt, only the E-W direction is taken into account, since Sorrells' theory only applies to the direction of the mean wind. In this case, the waveform is reproduced well - especially during the windy period - but Sorrells' method overestimates the tilt by a factor of about 2. This difference is likely related to the loss of information about the 2D complex structure of the pressure field in Sorrells' approximation. The difference in amplitude does not affect the efficiency of the decorrelation techniques (see \sect{decorel}), but it needs to be taken into account when estimating the ground properties from the comparison of pressure and tilt data.

\begin{figure}[h!]
\begin{center}
\includegraphics[scale=0.3]{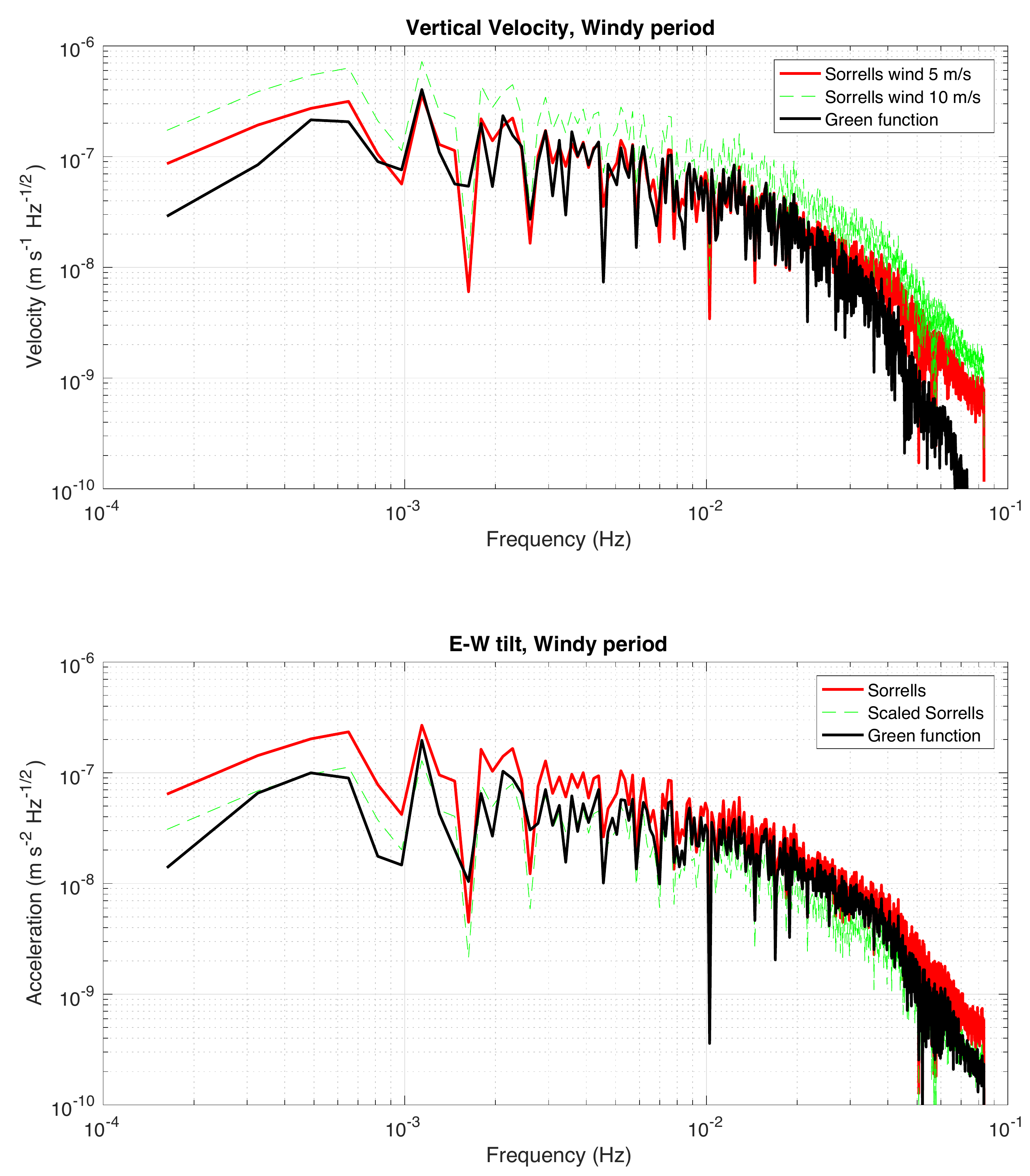}
\caption{(Upper figure) Comparison of the frequency content of the vertical velocity during the windiest period calculated using Green's function method (black) and the single-station Sorrells' approach with a wind speed of 5 m/s (solid red), and with a wind speed of 10 m/s (dashed green). (Lower figure) Comparison of the frequency content of the ground tilt in the E-W direction calculated using Green's function method (black), the single-station Sorrells approach (red), and the scaled Sorrells' results (dashed green).}
\label{fig:SorrelsGreensSpectra}
\end{center}
\end{figure}

The frequency content of the pressure noise signal predicted using Sorrells' single-station method and Green's function method can be seen in \fig{SorrelsGreensSpectra}. At low frequencies (up to 0.3 Hz), the vertical velocity spectra of Green's function calculations and Sorrells' single-station method with a wind speed of 5 m/s match very well. However, at higher frequencies, the vertical velocity spectrum from Green's function calculations drops off much more sharply than Sorrells' method calculations. Although this may be due to a numerical issue close to the Nyquist frequency, it might also be related to a loss of coherency of the pressure at short periods, leading to a smaller ground tilt than those generated by the transported pressure field assumed by Sorrells. Again, the amplitude is larger for Sorrells' single-station method with a wind speed of 10 m/s. The tilt spectra (\fig{SorrelsGreensSpectra}) show an excellent agreement between models across the entire frequency band. 

\newpage
\section{Correlations between the seismic and pressure signals}

\begin{figure}[h!]
\begin{center}
\includegraphics[scale=0.25]{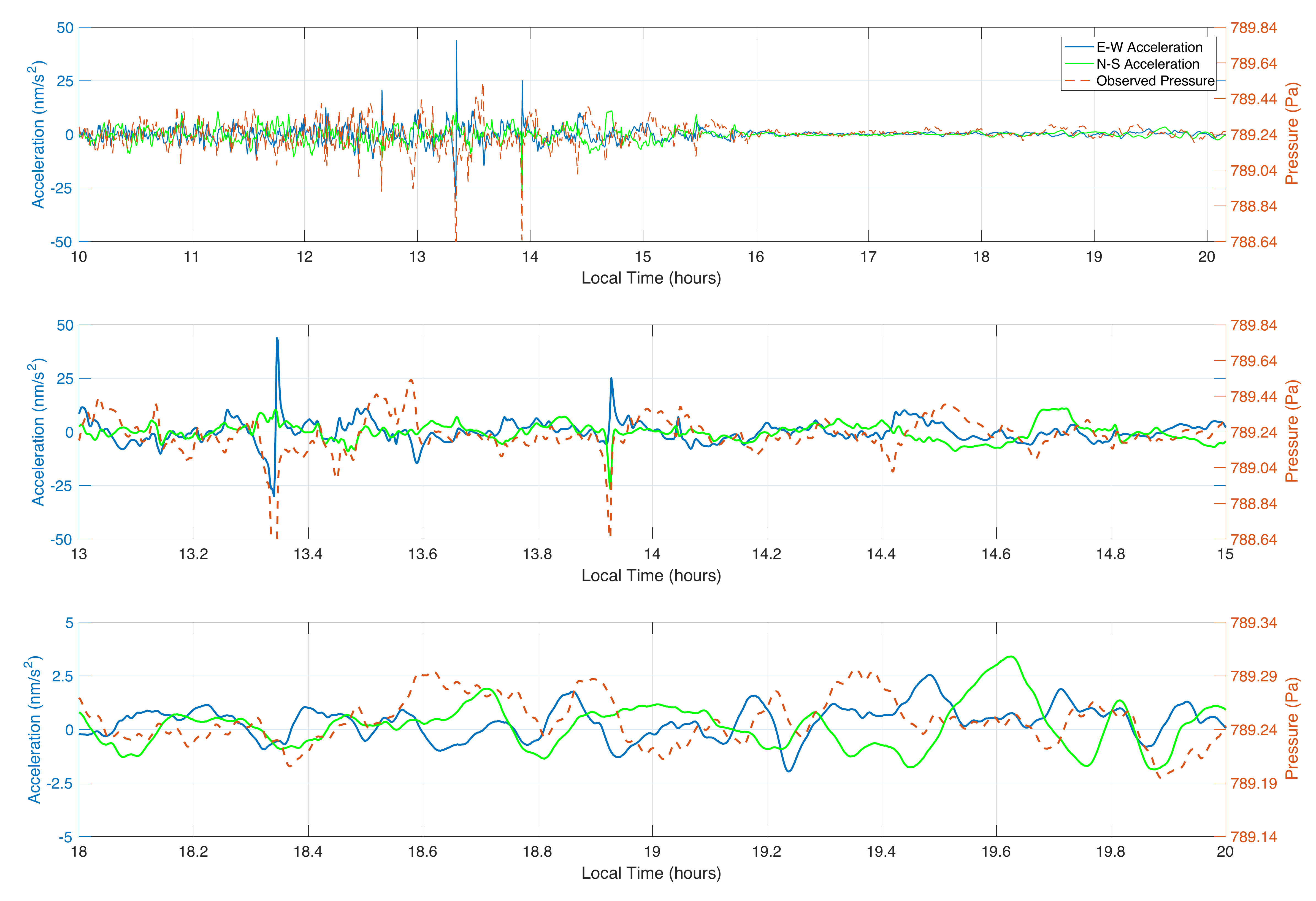}
\caption{The acceleration due to tilt in the E-W direction (solid blue) and in the N-S direction (solid green) as a function of time due to the pressure fluctuations calculated using Green's function method plotted on the same figure as the pressure observed at the location of the seismometer (dashed red). The top figure shows the signals over the full LES simulation, whereas the middle figure and lower figure show the windiest and calmest periods, respectively.}
\label{fig:PressureTilt3figs}
\end{center}
\end{figure}

The simultaneous tilt (Green's function method) and pressure signals are shown in \fig{PressureTilt3figs} for the entire simulated period, the windiest period and the calmest period. The distinctive signal of a dust devil can be seen in both the pressure and seismic data $\sim$13.3h and $\sim$13.9h. There is a sharp dip in local pressure that is coincident with a ``heartbeat'' seismic signature on one horizontal seismic axis and a large seismic signal on the other horizontal seismic axis. As the dust devil crosses (or passes close to) the seismometer, the ground tilts away from the negative pressure load of the vortex. The tilt rises from zero to some maximum value, which then switches sign as the load crosses the instrument and then declines back to zero. The component of tilt orthogonal to the direction of motion rises to a maximum value at close approach and declines (but is always of the same sign). The seismic signals of vortex features, first described in \cite{lorenz2015}, are studied in more detail in \cite{kenda2016}. 

%

\begin{figure}[h!]
\begin{center}
\includegraphics[scale=0.25]{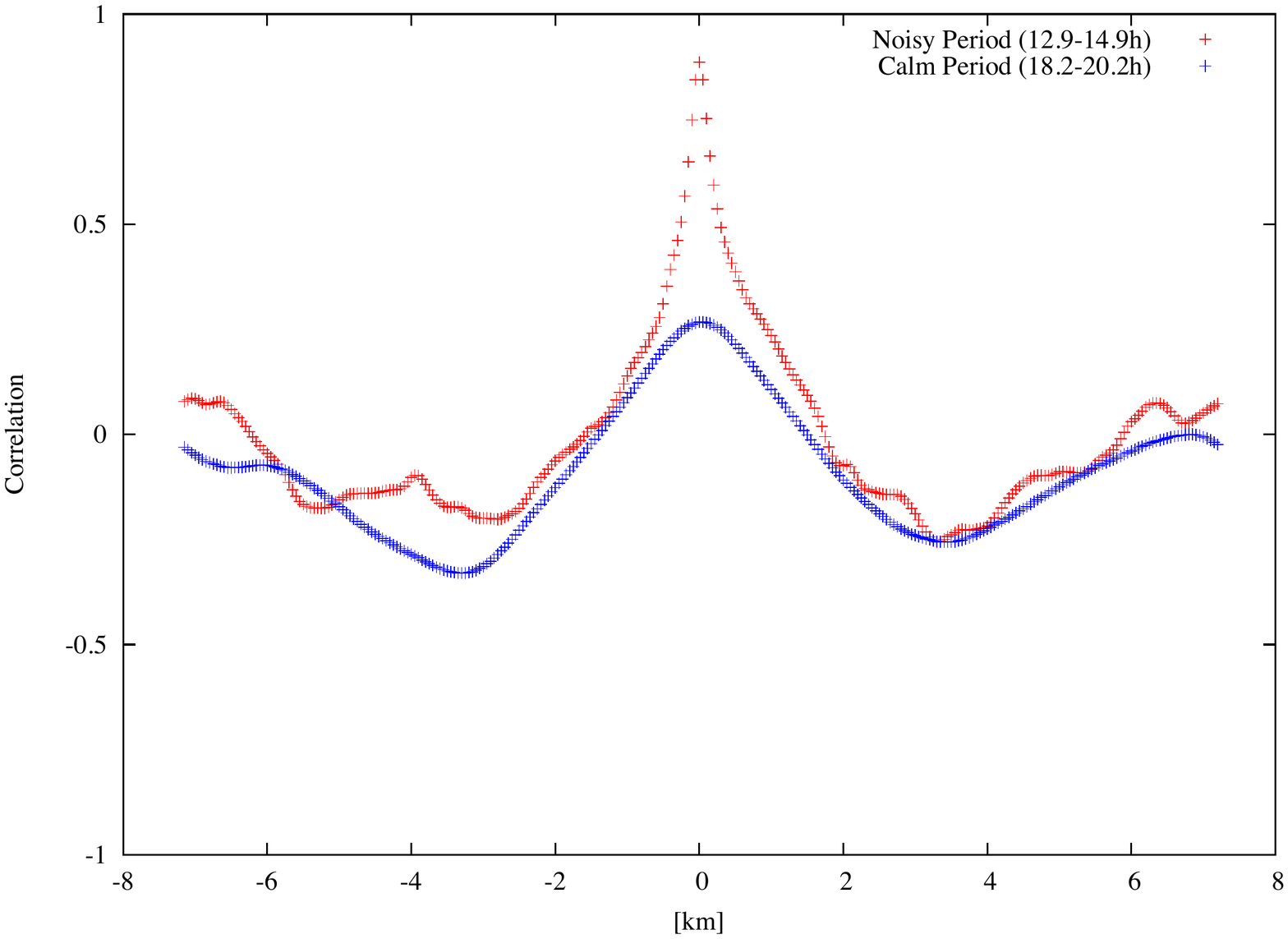}
\caption{Spatial correlation between the seismic pressure noise at the seismometer and the pressure signal at different distances from the seismometer along the wind direction (E-W direction). The positive and negative values correspond to the east and west directions, respectively. Correlations of both the windiest (red) and calmest (blue) periods are shown.}
\label{fig:FigureTK1}
\end{center}
\end{figure}

To better understand the sensitivity of the seismometers to the surrounding pressure fluctuations, the correlation between the seismic noise at the seismometer location and the pressure signal at a certain distance from the seismometer is calculated using the Pearson Correlation Coefficient as in \sect{pressurecorrelation}. The horizontal tilt towards the wind direction and the pressure signal every 50 m along the wind direction (\ie the highest resolution possible) are used. The evolution of the correlation between the two signals with distance is evaluated. As previously stated, two time windows are selected for the analyses; the windiest period (from 12.9h to 14.9h) and the calmest period (from 18.2h to 20.2h).  Note that before evaluating the correlation, the phase shift predicted from Sorrells' theory (see previous section and \eqn{sorrell} for details) was taken into account. As expected from the correlation of the pressure signals (\fig{PressurePearson}), the correlation of the seismic signal with pressure signal also decreases with distance (\fig{FigureTK1}).   

At $\sim \pm$3 km, the pressure field is significantly anti-correlated with the seismic noise. Then the correlation approaches zero for longer distances, where the seismic noise seems to be uncorrelated with the pressure field. Given that the correlation shows both positive and negative values, we can expect some periodic structure of km order around the seismometer. This is in good agreement with the LES results, and demonstrates clearly that seismic pressure noise reflects the atmospheric structure close to the seismometer. When the correlation of the windiest and calmest periods are compared, it can be seen that both show a large-scale structure of km order while, during the day-time, some smaller structure is superposed on-to the large-scale structure.  This can be understood as follows: during the calm period, which is mainly the night-time, the main source of tilt noise is the large-scale pressure variation. On the other hand, during the windiest period, which mainly corresponds to the day-time, the main source of noise is the turbulence associated with the convective cells which dominate at the smaller scale of $<$1km. In addition to this, it can be seen that the correlation between the seismic signal and the pressure signal is higher for the windiest period.  

\begin{figure}[h!]
\begin{center}
\includegraphics[scale=0.25]{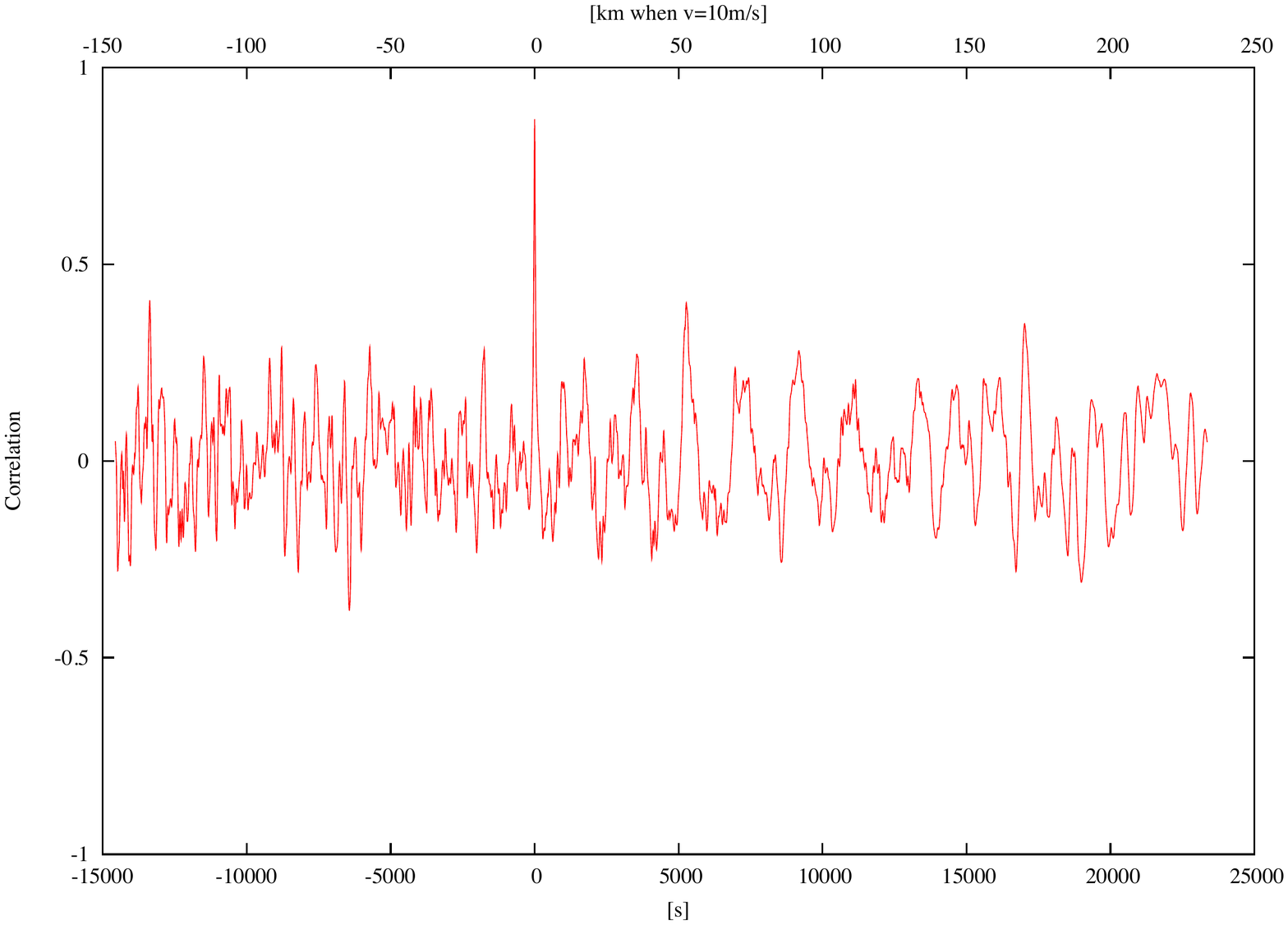}
\caption{Temporal correlation between the seismic pressure noise and the co-located pressure signal during the windiest period. The correlation is shown as a function of time shift between the two signals (lower x-axis). The time shift was converted into a spatial distance assuming the wind velocity of 10m/s (the background wind of the LES) giving also a spatial correlation (upper x-axis). }
\label{fig:FigureTK2}
\end{center}
\end{figure}

\begin{figure}[h!]
\begin{center}
\includegraphics[scale=0.25]{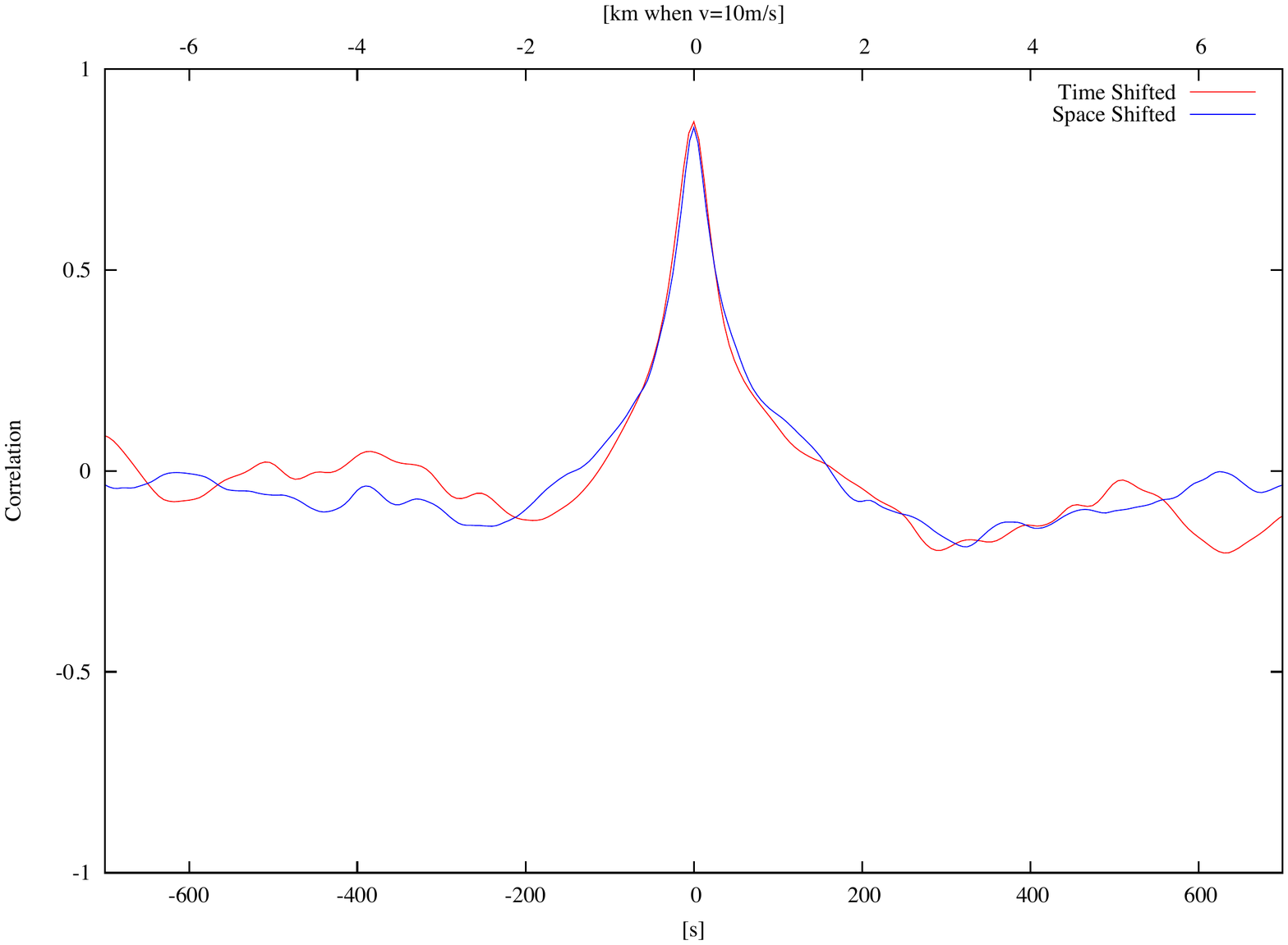}
\caption{Comparison of the correlation between the seismic pressure noise and the pressure signal calculated with two different methods.  In blue is the spatial correlation of the seismic pressure noise at the seismometer and the pressure signal at different distances from the seismometer that has also been converted into a temporal shift by assuming the wind velocity of 10 m/s (as in \fig{FigureTK1}).  In red is the temporal correlation of the seismic pressure noise and the co-located pressure signal that has also been converted into a spatial shift by assuming the wind velocity of 10m/s (as in \fig{FigureTK2}). The windiest period was used here, and the correlations along the wind direction are considered (E-W direction). }
\label{fig:FigureTK3}
\end{center}
\end{figure}

Sorrells' theory describes that the pressure fluctuations that generate the seismic noise are carried by the mean wind speed of the field, and this results in time variation of the seismic noise. This implies that the temporal and spatial variations of the seismic noise are related by wind speed. A simple way to demonstrate this is to compare the correlation in the spatial domain with the correlation in time domain. \Fig{FigureTK2} shows the correlation in the time domain of the observed pressure signal and the seismic signal. Note that the periodicity apparent in \fig{FigureTK2} is related to the periodic boundary conditions of the LES; the grid is 14.4 km wide, and the background wind is 10 m/s resulting in a periodicity of 1440 seconds. A 5000 second time-moving window was used to calculate the time evolution of the correlation over the entire data set. Assuming a mean wind velocity of 10 m/s (the background wind imposed in the LES), the time domain can be converted to the space domain, and this can be compared to what was obtained in the spatial correlation. \Fig{FigureTK3} shows this comparison, and it can be seen that the two correlations agree very well. This clearly shows that the noise source, or pressure fluctuations, are carried by the wind field. 



\newpage
\section{Decorrelation of the pressure signal}
\label{s:decorel}

Previous discussions demonstrate that the pressure variation around the seismometer is a significant source of noise on the seismometer. While such noise is not the primary noise source in the terrestrial environment, this will be the main source of noise on Mars due to the lack of oceanic noise and the soft regolith layer under the seismometer. Noise decorrelation with the pressure signal obtained with pressure sensor, part of InSight APSS (Auxiliary Payload Sensor Suite), will play an important role in the preliminary data processing of the SEIS data. In this section the decorrelation strategy is described and tested using the data provided in the previous sections.

The basic approach is to decorrelate the seismic signal that is coherent with the pressure signal using a FIR (finite impulse response) filter.  In the previous section it has been shown that the expected seismic noise is strongly correlated with the pressure signal, particularly for the local-scale atmospheric structure.  It is important to note that local pressure signal at the seismometer location showed a high correlation with the seismic signal, despite the fact that all the pressure fluctuations within the vicinity of the seismometer generate noise. This strongly supports the decorrelation strategy employed here and implies that a significant portion of the pressure noise should be able to be decorrelated on the seismic records. 


The first step of the decorrelation process is to tune the FIR filter through the least square fits of pressure signal to the seismic noise. We assume that the noise generated by the pressure noise can be expressed with a simple equation:

\begin{equation}
d_i= \sum_{j=-N}^{N} B_j P_{i+j}
\end{equation}

\noindent where $d_i$ is the pressure noise, $P_{i+j}$ is the pressure signal at time sample $i+j$ and $B_j$ is the FIR filter of  $2N+1$ coefficients. The FIR filter is defined by finding $B$ that minimizes the difference between $d_i$ and the seismic signal $s_i$ for the given time window. 

While pressure variations generate seismic noise, ground motions from seismic activity also result in pressure variations, { with a very well known relation $P = \rho_0  c_0  V$, where $P$, $\rho_0$, $c_0$ and $V$ are pressure, atmosphere density and vertical velocity, respectively \citep{Mikumo2009}. For a typical Martian atmosphere with density and sound speed of 0.02 $kg/m^3$ and 240 $m/s$, respectively, this provides a conversion factor from the ground velocity to the pressure of about 4.8 ${ \mu Bar \over \mu m / s}$. This is one order of magnitude smaller than the conversion factor obtained from \eqn{sorrell}, which provides a conversion factor of about 50 ${\mu Bar \over \mu m /s}$ for 10 m/s wind, assuming the values in \tbl{regolith}. This means that the pressure decorrelation, the coefficients of which will be based on the inverse of those noted above, from pressure to ground velocity, will not significantly affect the seismic signal, even if the pressure-converted wave is present in the pressure signal. This is a very different case from Earth, where the conversion factor of seismic waves is 70 times higher and is, therefore, larger than the one associated with the elastic deformation.  Nevertheless, and in order to not to influence the seismic data from Mars, a time window without any seismic signal will have to be chosen to define the FIR filter for future data, and the same filter will then be applied to the other time windows.

As there is no seismic event within our synthetic test data, the entire data set could be used to tune the FIR filter.} However, the approach that will be applied to the Martian data - using two different time windows - has been maintained here in order to test the stability of the FIR filter. Due to the change in the geometric configuration of the pressure field and the seismometer location, the transfer function may change and the stability of the filter will be an important factor for the decorrelation efficiency. 

\begin{figure}[h!]
\begin{center}
\includegraphics[scale=0.22]{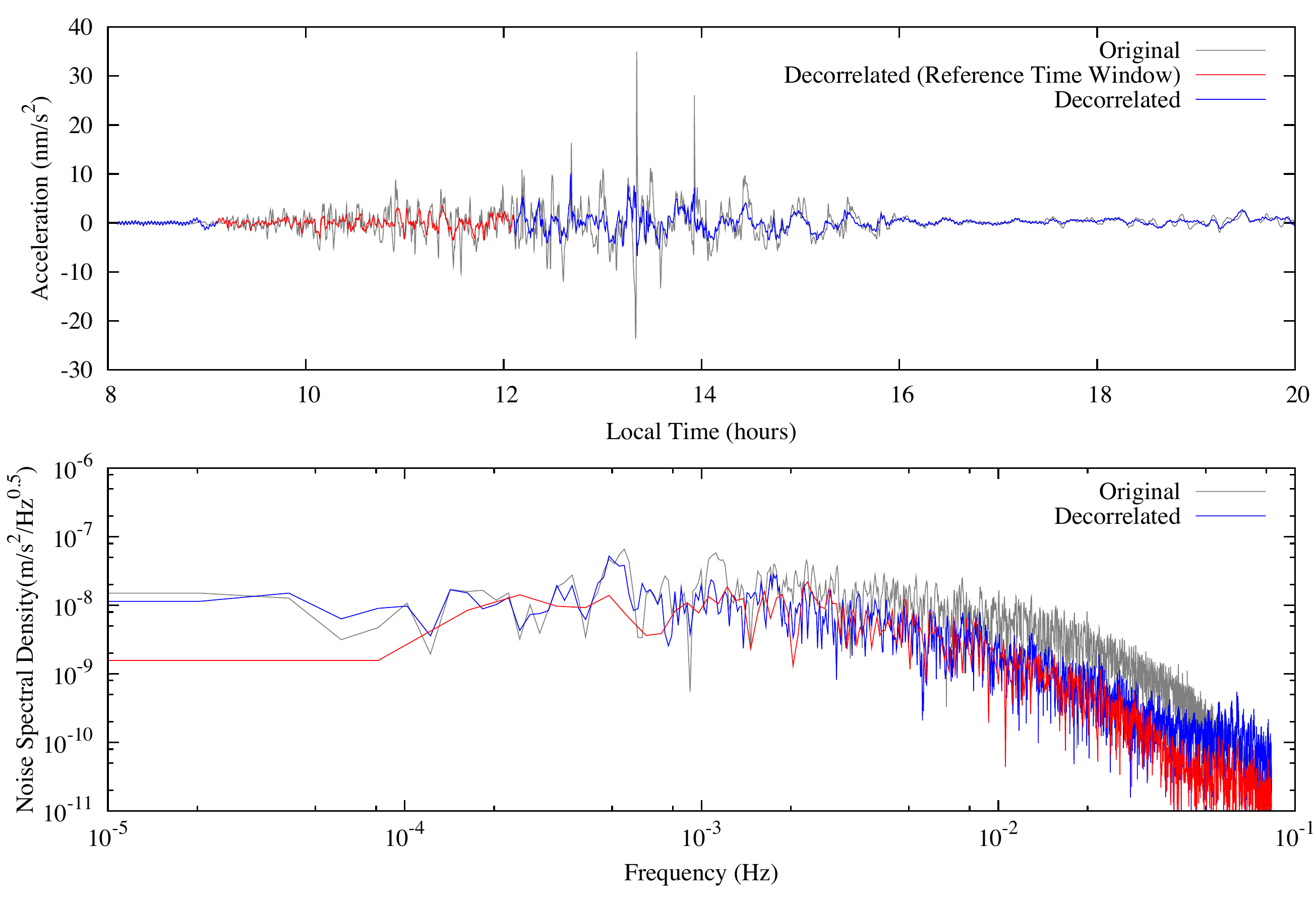}
\includegraphics[scale=0.22]{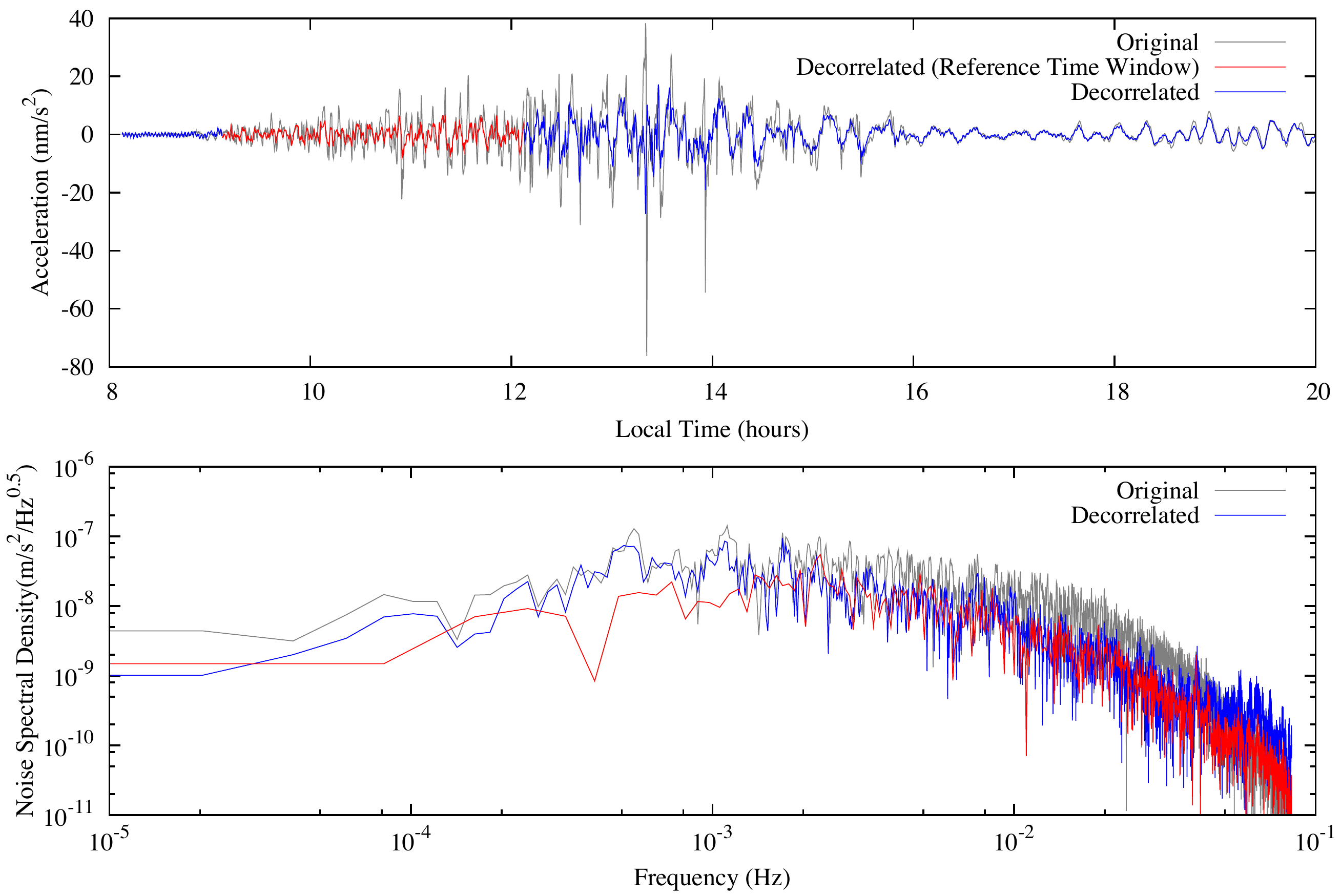}
\caption{Example of pressure noise decorrelation performed on (left) horizontal acceleration in the mean wind direction and (right) vertical velocity. The top panels show the data in the time domain and the bottom panels show the frequency domain. The original data is shown in grey; the decorrelated data within the reference time window are indicated in red, and the decorrelated data outside the reference time window are indicated in blue. The FIR filters used for the decorrelation were defined in the reference time window. The size of the filter used here is 101 coefficients for the horizontal acceleration and 141 coefficients for the vertical velocity. The FIR filter was applied both to the signal within (red) and outside (blue) the reference time window.}
\label{fig:FigureTK11}
\end{center}
\end{figure}

\begin{figure}[h!]
\begin{center}
\includegraphics[scale=0.23]{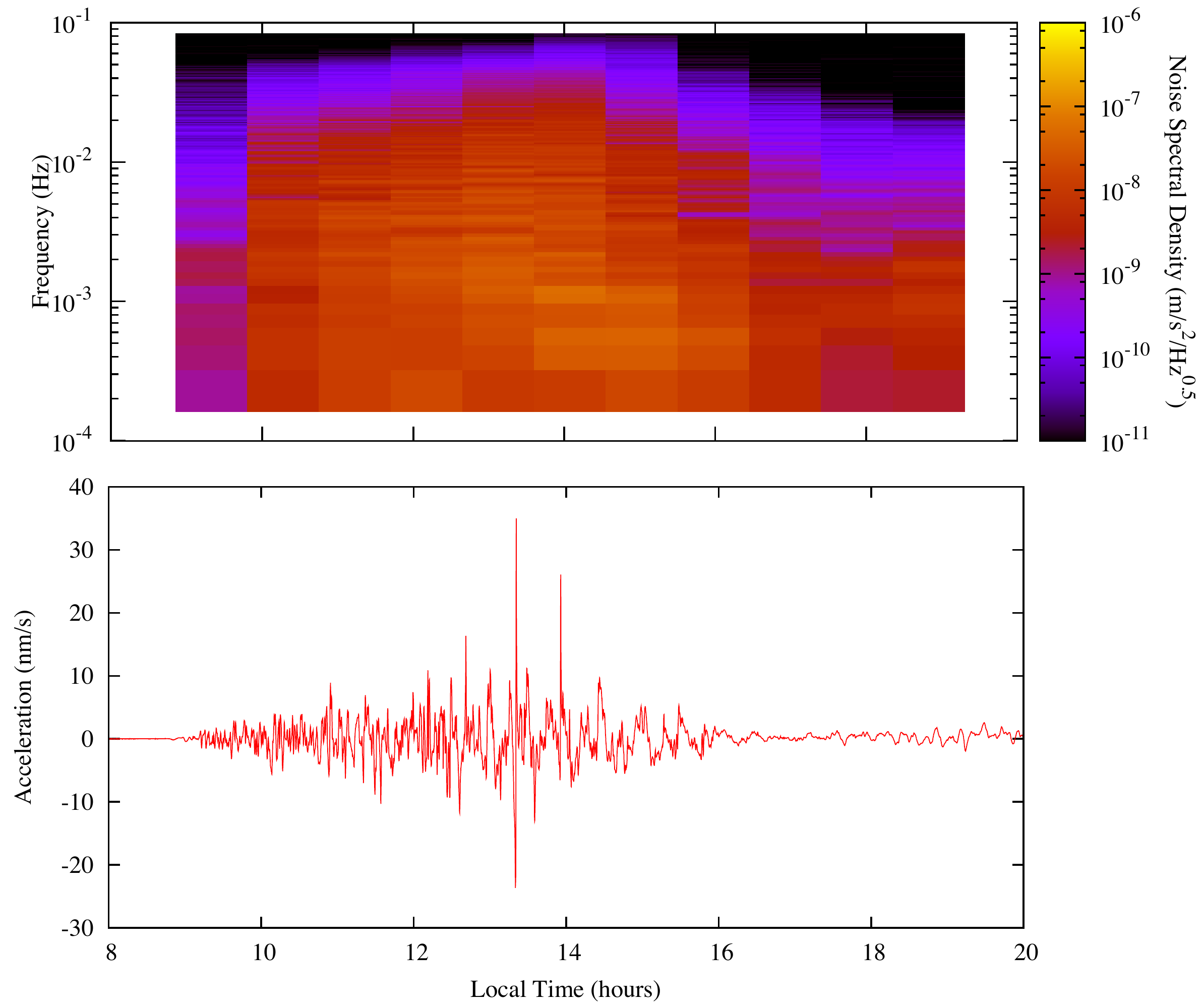}
\includegraphics[scale=0.23]{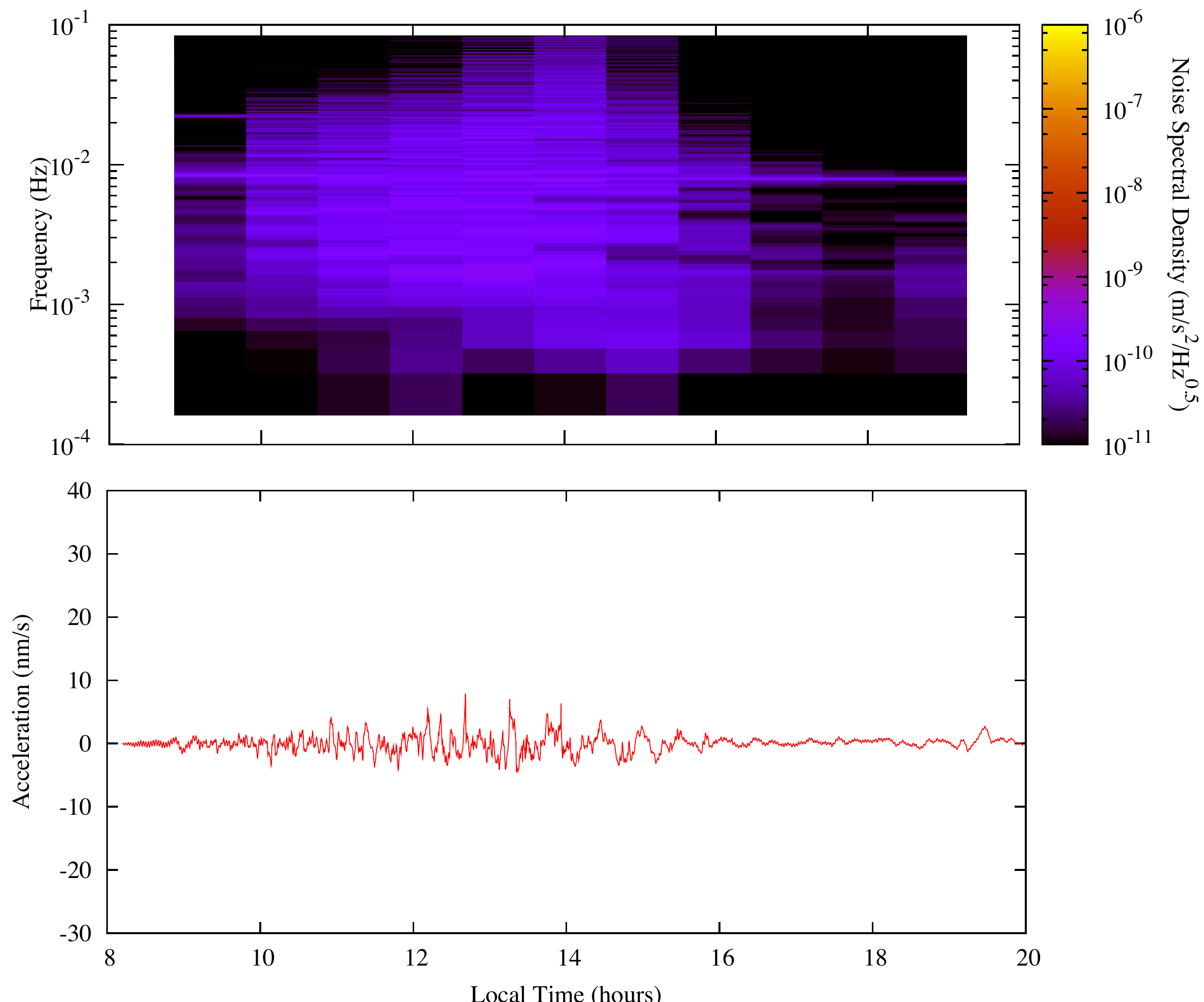}
\caption{(Left) Time series and spectrogram of horizontal acceleration on mean wind direction before the decorrelation. One-hour time window was used to construct the spectrogram. (Right) Time series and spectrogram of horizontal acceleration on mean wind direction after the decorrelation. The decorrelation was performed in the same manner as described in \fig{FigureTK11}. One-hour time window was used to construct the spectrogram.}
\label{fig:FigureTK21}
\end{center}
\end{figure}

\Fig{FigureTK11} shows an example of pressure noise decorrelation demonstrated for the horizontal tilt in the E-W direction (the wind direction). Here, the beginning of the windy period (9h-13h) is used as a reference time window to design a FIR filter, and this is applied to the rest of the time series. The FIR filter length of 1-1001 coefficients (i.e. 6-6000s) has been tested, and it was found that the FIR length of 80-101 coefficients is suitable for the decorrelation. The decorrelation efficiency has also been evaluated using the spectrograms. \Fig{FigureTK21} (left) shows time series and corresponding spectrogram of seismic noise before and after the decorrelation. The spectrograms consist of spectra calculated for a 1-hour time window. When the amplitude spectral density spectra before and after the decorrelation are compared, a significant improvement of noise level can be observed, especially in the 0.001-0.05 Hz bandwidth where the noise level is reduced by a factor of $\sim$5. 

\Fig{FigureTK11} (lower left) shows both the spectra of the decorrelated signal within the reference time window (red) and outside the reference time window (blue). We see that within the frequency band 0.001-0.05 Hz, the decorrelation outside the reference time window is as good as the decorrelation within the reference time window. However, the decorrelation adds some numerical noise at higher frequencies, especially outside the reference time window. This implies that there is some evolution of the transfer function for the pressure noise. 

Another implication from this test is that the noise decorrelation is more efficient for higher frequencies (0.01-0.05 Hz), and the efficiency decreases at lower frequencies. This is evident also in the time domain data where the long-period signal is still visible after the decorrelation. The time series plot (\fig{FigureTK11}, upper left) also shows that the decorrelation is more efficient at the noisy period compared to the calm period. These features can be understood from the higher correlation and shorter correlation length seen during the windiest period as seen in \fig{FigureTK1}. \Fig{FigureTK1} shows that during the windiest period, the seismic noise is highly correlated with the pressure field $\sim$1km around the seismometer, which generates noise at higher frequency or at shorter wavelength. On the other hand, during the calm period, the seismic noise is correlated with a larger scale pressure field of $\sim$4km, which will generate noise at lower frequency and longer wavelength. The correlation is lower compared to the noisy period and thus makes the decorrelation difficult. 

The same data processing was performed also for the horizontal tilt perpendicular to the wind direction. However, this was not as successful as it was for the tilt along the wind direction. This is consistent with what has been predicted by previous studies \citep{sorrells1971,sorrells1971b} where the ground will be tilted towards the wind direction. This implies that the wind direction data from the anemometer, another part of InSight APSS, will be an important source of information to align the seismic data and carry out an efficient pressure noise decorrelation.

\begin{figure}[h!]
\begin{center}
\includegraphics[scale=0.23]{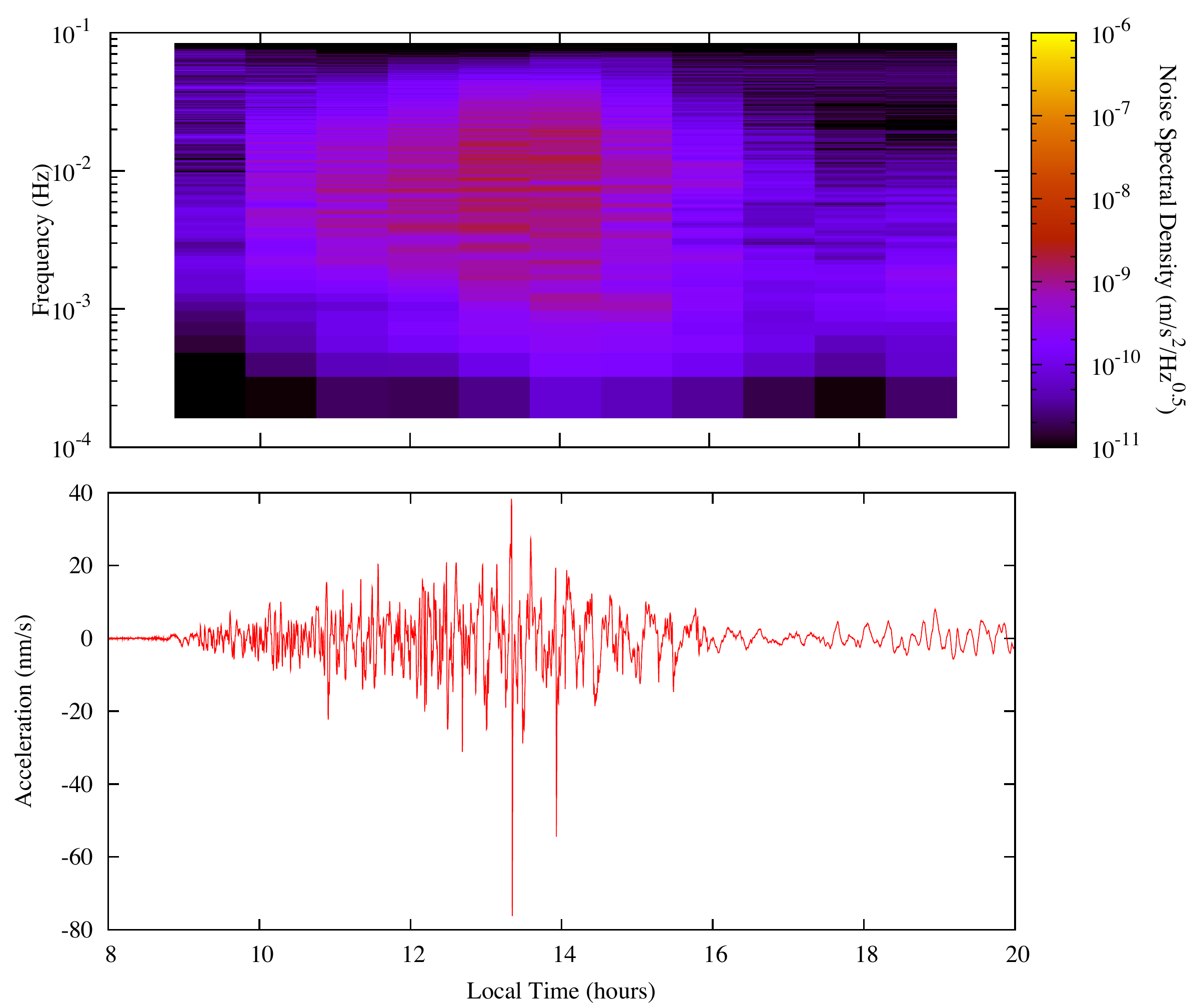}
\includegraphics[scale=0.23]{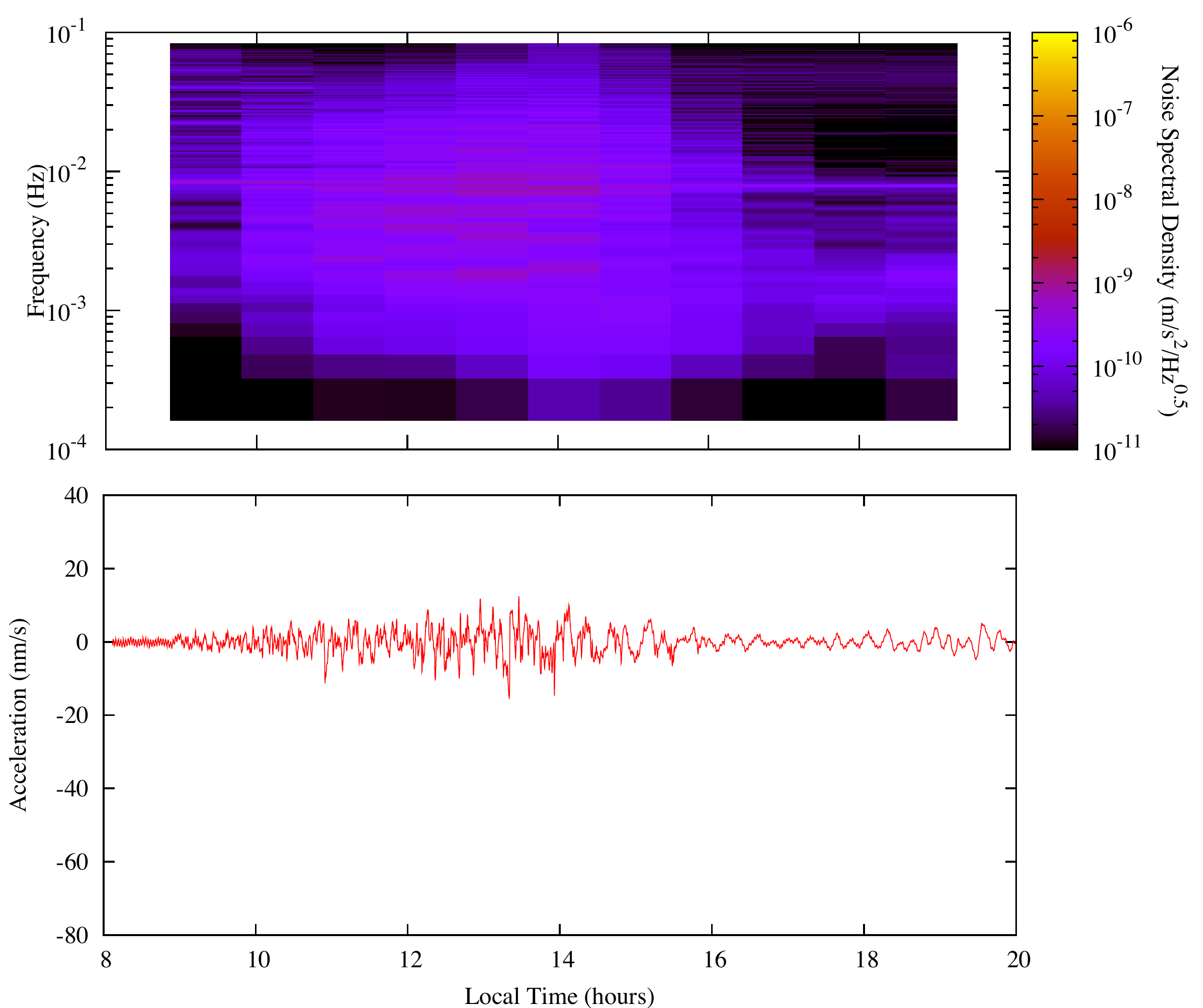}
\caption{(Left) Time series and spectrogram of vertical velocity before the decorrelation. One-hour time window was used to construct the spectrogram. (Right) Time series and spectrogram of vertical velocity after the decorrelation. The decorrelation was performed in the same manner as described in Figure TK11. One-hour time window was used to construct the spectrogram. }
\label{fig:FigureTK23}
\end{center}
\end{figure}

\Fig{FigureTK11} shows the results from the same decorrelation process with the vertical component.  As before, the spectrogram are also shown in \fig{FigureTK23}.  As discussed above, Sorrells' theory predicts that the pressure is proportional to the vertical velocity and, therefore, the vertical velocity was used instead of the vertical acceleration for the decorrelation. Though the noise level on the vertical component is significantly lower than that on the horizontal axis, a similar improvement in the noise level (factor of $\sim$5) was found for the vertical component. 

An advantage of using a time-evolving FIR filter for decorrelation is that the FIR filter can be efficiently retuned to decorrelate sporadic pressure-induced seismic sources. For example, in the specific case of dust devils (or convective vortices) these local, sporadic events have clearly identifiable signals; the ``heartbeat'' seismic signature, as described in the previous section. The transfer function between the vortices pressure signal and the seismic noise is likely to be different to the transfer between the background pressure signal and the seismic noise. Thus, as a time-evolving FIR filter is being used for decorrelation, it will be possible to retune the FIR filter to efficiently decorrelate the dust devil. Then, once the dust devil has passed, the FIR filter can be retuned to match the pressure field without the dust devil.  If, however, the transfer function of the different pressure signals is constant, there is not an issue; our aim is to decorrelate the pressure noise regardless of the source. 

The decorrelation strategy tested with the synthetic noise data obtained from the LES pressure field clearly shows that this strategy is efficient for the pressure noise decorrelation. It is, however, likely that the decorrelation will be less efficient when noises from multiple sources are superposed. This will be investigated in future tests that are expected before the InSight launch in 2018.

\section{Conclusions}

The atmospheric pressure fluctuations on Mars will induce an elastic response in the ground that will create a ground tilt, detectable as a seismic signal on SEIS. This ground tilt due to atmospheric pressure variations is anticipated to be a major seismic signal on the SEIS instrument \citep{mimoun2016}. It is planned to reduce the atmospheric seismic signal by making use of a pressure sensor that will be part of the InSight APSS (Auxiliary Payload Sensor Suite). Decorrelation techniques will be used to  remove the pressure signal from the seismic signal. The pressure sensor will be on the InSight lander and, thus, almost collocated with the seismometer. 

Here we use Large Eddy Simulations (LES) of the wind and surface pressure at the InSight landing site combined with ground deformation models to investigate the atmospheric pressure signals on SEIS. The seismic pressure noise is calculated using the LES-predicted surface pressure at the InSight landing site and a Green's function approach. The horizontal acceleration as a result of the ground tilt in the E-W and N-S directions due to the LES pressure field is found to be typically $\sim$2 - 40 nm/s$^2$ in amplitude, whereas the direct horizontal acceleration is two orders of magnitude smaller than the contribution to the acceleration from the ground tilt and is thus negligible in comparison. The vertical accelerations are found to be $\sim$0.1 - 6 nm/s$^2$ in amplitude.

The Green's function approach to seismic simulations are validated via a detailed comparison with two other independent methods: a spectral approach using the entire pressure field \citep[as in][]{kenda2016}, and a single-station approach based on Sorrells' theory \citep{sorrells1971,sorrells1971b} and using only the co-located seismic and pressure measurements.  These three models all assume the same ground properties and use a half-space approximation. This ground model gives a worst-case estimate for the seismic noise, as the presence at some (shallow) depth of a harder layer would significantly reduce quasi-static displacement and tilt effects.

The investigations of the correlation between the pressure signal at the center of the LES field with the pressure signal in the vicinity have shown that under calm conditions, a single-pressure measurement is representative of the large-scale pressure field (to a distance of several kilometers), particularly in the prevailing wind direction. During windy conditions, however, small-scale turbulence results in a reduced correlation between the pressure signals, and the single-pressure measurement becomes less representative of the pressure field.  Nonetheless, the correlation between the seismic signal and the pressure signal is found to be higher for the windiest period despite the fact that a single-pressure measurement is less representative of the entire pressure field during windy conditions.  This is because the seismic pressure noise reflects the atmospheric structure close to the seismometer, particularly during windy periods; during the calmer periods (mainly the night-time) the main source of pressure noise is the large-scale ($>$1km) pressure variation, but during the windy periods (mainly the day-time), the main source of pressure noise is the turbulence excited by the convective cells which dominate at the smaller scale of $<$1km.  It has also been confirmed that the noise source, or pressure fluctuations, are carried by the wind velocity field as described by Sorrels \citep{sorrells1971,sorrells1971b}.

Finally, the InSight decorrelation strategy tested with the synthetic noise data obtained from a LES pressure field clearly shows that this strategy is efficient for the pressure noise decorrelation. Indeed, a reduction by a factor of $\sim$5 is observed in the horizontal tilt noise (in the wind direction) and the vertical noise in the 0.001-0.05 Hz bandwidth. This suggests that low, long-period noise levels can be envisaged, especially if bedrock layers are expected at a depth of several meters. It is, however, likely that the decorrelation will be less efficient when noises from other sources (e.g. magnetic, thermal and instrument self noise) are superposed. This will be investigated in future tests that are expected before the InSight launch in 2018. 

\section{Acknowledgments}

{ This work has been supported by CNES and by ANR SEISMARS, including post-doctoral support provided to N. Murdoch and to T. Kawamura. B. Kenda acknowledges the support of the ED560 STEP'UP and of the NASA InSight project for his PhD support. This paper is InSight contribution 23.}

\bibliographystyle{aps-nameyear}      


\end{document}